\documentclass[13pt]{article}

\usepackage{url}
\DeclareMathAlphabet{\mathcal}{OMS}{cmsy}{m}{n}
\usepackage{subcaption}
\usepackage{amssymb,amsmath,amsfonts}
\usepackage{mathrsfs}
\usepackage{amsthm}

\usepackage{enumitem}
\usepackage{esvect}
\usepackage{tikz}
\usetikzlibrary{circuits}

\usetikzlibrary{intersections} 

\usetikzlibrary{scopes, arrows, fadings, patterns}

\usetikzlibrary{%
	decorations.pathreplacing,%
	decorations.pathmorphing%
}

\usetikzlibrary{positioning}

\usepackage{pgfplots}

\newcommand{\Eb}{{\mathbb{E}}}
\newcommand{\Rb}{{\mathbb{R}}}
\newcommand{\Cb}{{\mathbb{C}}}
\newcommand{\Ib}{{\mathbb{I}}}
\newcommand{\Zb}{{\mathbb{Z}}}

\newcommand{\Es}{{\mathscr{E}}}

\newcommand{\Ec}{{\mathcal{E}}}

\newcommand{\Fc}{{\mathcal{F}}}

\newcommand{\Ac}{{\mathcal{A}}}

\newcommand{\Ic}{{\mathcal{I}}}
\newcommand{\Jc}{{\mathcal{J}}}

\newcommand{\Uc}{{\mathcal{U}}}

\newcommand{\Sc}{{\mathcal{S}}}

\newcommand{\Mc}{{\mathcal{M}}}

\newcommand{\Wc}{{\mathcal{W}}}
\newcommand{\Lc}{{\mathcal{L}}}

\newcommand{\ra}{{\rightarrow}}
\newcommand{\ift}{{\infty}}
\newcommand{\ls}{\text{ls}}

\DeclareMathOperator{\rs}{{rowspan}}

\DeclareMathOperator{\diag}{{diag}}
\DeclareMathOperator{\supp}{supp}
\DeclareMathOperator{\tr}{tr}

\newtheorem{proposition}{\textbf{Proposition}}
\newtheorem{lemma}{\textbf{Lemma}}
\newtheorem{theorem}{\textbf{Theorem}}
\newtheorem{remark}{\textbf{Remark}}
\newtheorem{assumption}{\textbf{Assumption}}
\newtheorem{corollary}{\textbf{Corollary}}

\newtheorem{definition}{\textbf{Definition}}

\begin{document}

\title{\bf Efficient Secure State Estimation against\\
	 Sparse Integrity Attack for \\
	 System with Non-derogatory Dynamics}

%\author[1]{Zishuo Li}
%
%\author[2]{Yilin Mo*}

%\authormark{Zishuo Li and Yilin Mo}
%
%
%\address[1]{\orgdiv{Department of Automation and BNRist}, \orgname{Tsinghua University}, \orgaddress{\state{Beijing}, \country{China}}}
%
%\address[2]{\orgdiv{Department of Automation and BNRist}, \orgname{Tsinghua University}, \orgaddress{\state{Beijing}, \country{China}}}
%
%
%\fundingInfo{This work is supported by the National Key Research and Development Program of China under Grant 2018AAA0101601.}

%\jnlcitation{\cname{%
%\author{Williams K.}, 
%\author{B. Hoskins}, 
%\author{R. Lee}, 
%\author{G. Masato}, and 
%\author{T. Woollings}} (\cyear{2016}), 
%\ctitle{A regime analysis of Atlantic winter jet variability applied to evaluate HadGEM3-GC2}, \cjournal{Q.J.R. Meteorol. Soc.}, \cvol{2017;00:1--6}.}
\author{Zishuo~Li and Yilin~Mo
	% <-this % stops a space
	\thanks{Zishuo Li and Yilin Mo are with the Department
		of Automation and BNRist, Tsinghua University, Beijing, China, e-mail: \texttt{lizs19@mails.tsinghua.edu.cn, ylmo@mail.tsinghua.edu.cn}. }
	\thanks{This work is supported by the National Key Research and Development Program of China under Grant 2018AAA0101601.} }

\date{}

\maketitle

\textbf{Abstract: }{We consider the problem of estimating the state of a time-invariant linear Gaussian system in the presence of integrity attacks. The attacker can compromise $p$ out of $m$ sensors, the set of which is fixed over time and unknown to the system operator, and manipulate the measurements arbitrarily. Under the assumption that all the unstable eigenvalues of system matrix $A$ have geometric multiplicity 1 (unstable part of $A$ is non-derogatory), we propose a secure estimation scheme that is resilient to integrity attack as long as the system is $2p$-sparse detectable, which is proved to be the fundamental limit of secure dynamic estimation. In the absence of attack, the proposed estimation coincides with Kalman estimation with a certain probability that can be adjusted to trade-off between performance with and without attack. Furthermore, the detectability condition checking in the designing phase and the estimation computing in the online operating phase are both computationally efficient. A numerical example is provided to corroborate the results and illustrate the performance of the proposed estimator.}

\textbf{Keywords: }{secure estimation, sparse integrity attack, Kalman filter, sparse observability}

%\footnotetext{\textbf{Abbreviations:} ANA, anti-nuclear antibodies; APC, antigen-presenting cells; IRF, interferon regulatory factor}

\section{Introduction}

As the confluence of sensors, platforms, and networks increases, the already widespread applications of Cyber-Physical System (CPS) and Internet of Things (IoT) are expected to continue to emerge and expand \cite{2018DHS_report}.
They play an increasingly important role in critical infrastructures and everyday life, while the cyber-security risks and attack surfaces are also increasing \cite{cardenas2009challenges}.
However, CPS is vulnerable to a variety of cyber attacks since it usually relies on remote sensing devices, communication channels, and spatially distributed processors, which are prone to failures when exposed to unintentional faults and malicious attacks.
Failure of CPS may cause severe damage to industrial infrastructures, economic order, and environmental systems, e.g., the Stuxnet launched on Iran’s nuclear facilities\cite{STUXNET}, power blackouts in Ukraine \cite{Ukraine_Blackout}, North America and Europe \cite{2003_blackout}, etc. The research community has recognized the importance of CPS security, especially the design of secure detection, estimation, and control strategy\cite{cardenas2009challenges}. 

Recently, substantial research efforts have been devoted to secure state estimation against various types of attacks, such as deception (integrity) attacks \cite{mo2016tac, Pajic2017cdc} and
denial-of-service (DoS) attacks \cite{xiaoqiang2018tac,yangguanghong2021tac}. 
The integrity attacks focus on destructing system data integrity by stealthily manipulating
the transmitted data, whereas DoS attacks jeopardize the availability of data resources by blocking the communication channels. A review of the secure estimation against various attacks is referred to \cite{ding2021surveyTSMC}.
This paper focuses on secure estimation against sparse integrity attack where an unknown subset of sensors is compromised by the adversary. The measurements from those corrupted sensors can be manipulated arbitrarily by the adversary.
In order to identify the sparse malicious sensors and mitigate the impact of manipulated measurements, the main research paths include error correction approach based on compressed sensing and switch estimation approach based on fault identification.
The error correction approach usually takes measurements in a finite time window and adopts a sparsity-inducing optimization to handle the outliers. For example, minimizing the $\ell_0$ norm or its convex relaxation $\ell_1$ norm for lower computational complexity \cite{FawziTAC2014,Pajic2017TAC}. Fawzi et al. \cite{FawziTAC2014} derive the fundamental limit for state reconstruction in the absence of noise and increase the number of correctable errors by state feedback. This result is further generalized to the scenario where the set of attacked nodes can change over time in \cite{chang2018auto}. For the scenario with bounded noise, Pajic et al. \cite{Pajic2017TAC} provides rigorous analytic bounds on the estimation errors
for $\ell_0$ and $\ell_1$-based estimation procedures. Similarly, Shoukry and Tabuada \cite{ShoukryTAC2016} adopt a $2$-norm batch optimization approach for state estimation and a customized gradient descent algorithm to solve it efficiently.
These works provide fundamental limits for estimation against integrity attack, i.e., proves that $2p$-sparse observability is necessary for secure state recovery against $p$ compromised sensors. 
However, the sensory data out of the window are discarded in the finite time window approach, which may cause performance degradation and estimation delay.

Another solution is the switch estimator\cite{HespanhaACC2015, Shoukry2017, Mishra2017TCNS,yorie_TAC2018,yangguanghong2018tac,luTAC2019} where the system operator switches between multiple estimate candidates\cite{HespanhaACC2015,yorie_TAC2018} or sensor subset combinations \cite{Shoukry2017, Mishra2017TCNS,yangguanghong2018tac,luTAC2019} based on the evaluation of their reliability by consistency checking or malicious detection algorithms. However, the combinatorial nature of candidate estimates or sensor combinations poses challenges for storage or computation capability, and various solutions are proposed. Shoukry et al.~\cite{Shoukry2017, Mishra2017TCNS} aim at reducing searching complexity by Satisfiability Modulo Theory, and \cite{luTAC2019} reduces the number of candidates with the help of a set cover approach.
In view of the computational problems, Liu et al. \cite{liuxinghua-TAC2020} propose a secure estimation scheme based on decomposing Kalman filters into local estimators whose weighted sum recovers the Kalman estimate with a certain probability in the absence of attack. The local estimates are fused securely by a quadratic programming problem with an $\ell_1$ term to handle the sparse outliers. However, in the designing phase, the sufficient condition for estimation resiliency is computationally hard to validate. Moreover, the sufficient condition has a gap from $2p$-sparse observability, which is the fundamental limit \cite{ShoukryTAC2016} for state reconstruction.
Similar to \cite{liuxinghua-TAC2020}, other results in the literature\cite{FawziTAC2014,sandberg_TAC2014} also impose more restrictive conditions than $2p$-sparse observability and are NP-hard to validate. Besides the research paths aforementioned, cost function-based adaptive observers provide resilient estimation for scenarios with time-varying set of malicious sensors\cite{mitra2018CISS} and unknown inputs \cite{wu2018TCYB}. Arpan and Urbashi\cite{mitra2018CISS} propose a filtering and learning algorithm with adaptive gain, minimizing the cost function based on estimation error in normal operation and under attack on various sensor subsets. 
For the scenario with both unknown inputs and corrupted sensors, Wu et al. \cite{wu2018TCYB} propose a projected sliding-mode
observer-based estimation algorithm to minimize the reconstruction error. However, the convergence of the descent algorithm used to minimize cost function does not necessarily imply the estimation error is bounded, which means secure guarantee is not easily obtained in these approaches.

In this paper, we focus on LTI system with Gaussian noise and intend to propose an estimation scheme that is secure (has bounded estimation error covariance) to $p$ corrupted senors as long as the system is $2p$-sparse detectable, under the assumption that the stable part of $A$ is non-derogatory. This achieves the fundamental limit for dynamic state estimation since it is proved that if the system is not $2p$-sparse detectable, there is no secure estimator \cite{yorie_TAC2018}. Moreover, by introducing a non-derogatory assumption, the sparse detectability index can be computed computational easily. For general system matrix $A$ with geometric multiplicity of unstable eigenvalues larger than 1, it has been proved that computing sparse observability is an NP-hard problem\cite{mao2021computational}, and there is no computational efficient solution unless $\rm P=NP$. 
Therefore, our proposed scheme reduces the computational complexity significantly when possible and achieves the fundamental limit of secure dynamic estimation.
Preliminary versions of some of the results have been presented in \cite{zishuo2021cdc_arxiv}. This paper is significantly expanded from the previous work in the following points:
\begin{itemize}
	\item The estimator proposed in \cite{zishuo2021cdc_arxiv} is secure if the system is $2p$-sparse observable. We design a novel estimator to secure the stable states in this paper, and the condition is relaxed to $2p$-sparse detectable and thus achieves the fundamental limit of secure dynamic estimation.
	\item This paper quantifies the probability of recovering the Kalman estimation in the absence of attack and the estimation error upper bound under attack, which is not provided in \cite{zishuo2021cdc_arxiv}. Moreover, all the proofs of results in this paper are provided.
	\item The system of this paper is more general than \cite{liuxinghua-TAC2020,zishuo2021cdc_arxiv} since control input $u(k)$ is considered.
\end{itemize}

In summary, in this paper, we propose a secure dynamic estimation scheme for linear Gaussian systems, and it has the following merits:
%prove that the computation complexity of secure dynamic estimation problem with noise is 
%In this paper, we prove that the computation complexity can be reduced significantly under the assumption that all the eigenvalues of system matrix $A$ have geometric multiplicity 1.
%The secure state estimation design is improved upon the previous work \cite{liuxinghua-TAC2020} and has the following merits:
\begin{itemize}
	\item In the presence of $p$ compromised sensors, the proposed estimation is secure if the system is $2p$-sparse detectable, which achieves the fundamental limit of secure dynamic state estimation.
	\item In the absence of attack, the proposed estimation coincides with Kalman estimation with certain probability, which can be adjusted to trade-off the performance with and without attack.
	\item During the designing phase, the sparse detectability index can be computed with low complexity. Moreover, during the algorithm operating phase, the proposed estimation is formulated as the solution of a convex optimization problem based on LASSO \cite{LASSOTibshirani}, which can be computed efficiently.
\end{itemize}

\textit{Organization:} We introduce the problem formulation and preliminary results in Section \ref{sec:problem}. The main results are provided in Section \ref{sec:main_result} and collaborated by numerical simulation in Section \ref{sec:sim}. Section \ref{sec:conclusion} finally concludes the paper.

\textit{Notations:}
Cardinality of a set $\Sc$ is denoted as $|\Sc|$. $A{'}$ represents conjugate transpose of matrix $A$.
%The determinant of a matrix is represented by $\det(\cdot)$. 
Diagonal matrix with diagonal elements $A_1,\cdots,A_k$ is denoted as $\text{diag}(A_1,\cdots,A_k)$.
Denote the span of row vectors of matrix $A$ as $\rs(A)$.
The trace of matrix $A$ is represented as $\tr(A)$.
All-one vector with size $m\times 1$ is denoted as $\mathbf{1}_{m}$. $I_n$ is the identity matrix with size $n\times n$. 
$\Cb^{m\times n}$ ($\Rb^{m\times n}$) represents the set of complex (real) matrices with $m$ rows and $n$ columns.
The $i$-th entry of a vector $x$ is represented by $x_i$ or $[x]_i$. $\|\cdot\|_q$ represents the vector $q$-norm or (induced) matrix $q$-norm which is clear according to the context.

\section{Problem Formulation and Preliminary Results}\label{sec:problem}	
\subsection{Secure dynamic state estimation}
In this paper, we consider the linear time-invariant system with Gaussian noise:
\begin{align}
x(k+1)&=A x(k)+Bu(k)+w(k) , \label{eq:system} \\
y(k)&=C x(k)+v(k)+a(k) ,\label{eq:y_i_def}
\end{align}
where $x(k) \in \mathbb{R}^{n}$ is the system state, $w(k) \sim {N}(0, Q)$ and $v(k) \sim {N}(0, R)$ are i.i.d. Gaussian process noise and measurement noise with zero mean and covariance matrix $Q$ and $R$.  
Vector $u(k)\in \mathbb{R}^{d}$ is the external input.  
The vector $y(k)\in \mathbb{R}^{m}$ is the collection of measurement from all $m$ sensors, and $i$-th entry $y_i(k)$ is the measurement from sensor $i$.
The vector $a(k)$ denotes the bias injected by an adversary and $a_i(k)$ is the attack on sensor $i$. Define $$z(k)=C x(k)+v(k)$$ as the true measurements without the attack.
The initial state $x(0) \sim {N}(0, \Sigma)$ is assumed to be zero mean Gaussian and is independent from the process noise $\{w(k)\}$.

The secure dynamic estimation problem aims at recovering system state $x(k)$ at every time $k$ based on all historical observations and inputs $\{y(t),u(t) | 0\leq t\leq k\}$, where $y(k)$ has been partly manipulated by the malicious attacker.
It is conventional in the literature \cite{FawziTAC2014,Shoukry2017} that the attacker can only compromise a fixed subset of sensors with known maximum cardinality. 
Denote the index set of all sensors as $\Jc \triangleq\{1,2, \ldots, m\}$. 
For any index set $\Ic \subseteq \Jc,$ define the complement set to be $\Ic^{c} \triangleq$ $\Jc \backslash \Ic$. 
Define the support of vector $a\in\Rb^{n}$ as $\supp(a)\triangleq \left\{i| 1\leq i\leq n , a_i\neq0 \right\}$ where $a_i$ is the $i$-th entry of vector $a$.
We have the following assumptions on the malicious adversary. 

\begin{definition}[Sparse Attack]\label{def:attack}
	The attack called a $(p, m)$-sparse attack if the vector sequence $a(k)$ satisfy that, there exists a time invariant index set $\Ic\subseteq \Jc $ with $|\Ic| = p$ such that $\bigcup_{k=1}^{\infty} \supp\left\{a(k)\right\} = \Ic$.
\end{definition}

Closely related to the sparse attack, we introduce the notion of sparse observability (detectability) that characterizes the system observability (detectability) in the presence of attack.

\begin{definition}[Sparse observable / detectable]\label{df:sparse_obs}
	The sparse observability (detectability) index of system \eqref{eq:system}-\eqref{eq:y_i_def} is the largest integer $s$ such that system\footnote{The matrix $C_{\Jc\setminus\Ic}$ represents the matrix composed of rows of $C$ with row index in $\Jc\setminus\Ic$.} $(A,C_{\Jc\setminus\Ic})$ is observable (detectable) for any set of sensors $\Ic\subset\Jc$ with cardinality $|\Ic| = s$. When the sparse observability (detectability) index is $s$, we say that the system with pair $(A,C)$ is $s$-sparse observable (detectable).
\end{definition}
\begin{remark}
	Since one can do Kalman decomposition for uncontrollable or unobservable systems and obtain a minimal realization which is both controllable and observable, we assume the system \eqref{eq:system}-\eqref{eq:y_i_def} is controllable and observable without loss of generality. Thus, the sparse observability (detectability) index is a non-negative interger.
\end{remark}
Define $y(k_1:k_2)$ as the sequence $\{y(k_1),y(k_1+1),\cdots,y(k_2)\}$. Similar notation is also applied on $z(k),u(k)$.	
An estimator is an infinite sequence of mappings $g=\{g_k\}_{k=1}^{\ift}$ where $g_k$ is a mapping from all the historical outputs and inputs to a state estimation at time $k$:
$$g_k\left(y(0:k),u(0:k)\right)=\hat{x}(k).$$
It is written as $g_k(y,u)=\hat{x}(k)$ for notation simplicity.
For linear Gaussian noise system, the estimation is secure if the estimation error covariance is bounded by a constant term irrelevant to the attack. 

\begin{definition}[Secure estimator]\label{def:resi}
	
	Define the estimation difference introduced by attack as
	\begin{align*}
	q_k\triangleq &g_k\left(z,u\right)-g_k\left(y,u\right) \\
	=&g_k\left(z,u\right)-g_k\left(z+a,u\right) .
	\end{align*}
	The estimator is said to be secure against $(p, m)$-sparse attack if the following holds:
	%there exists a constant $q$ such that  
	$$\sup_{k\in\Zb^+} \Eb \left[\tr\left(q_k q_k'\right)\right] < \ift ,$$
	where $\Eb$ is the expectation with respect to the probability measure generated by the Gaussian noise $\{w(k)\}$ and $\{v(k)\}$. 
\end{definition}
If all sensors are benign, i.e., $a(k)=\mathbf{0}$ for all $k$, the optimal state estimator is the classical Kalman filter:
\begin{align*}
\hat{x}(k)&=\hat{x}(k | k-1)+K(k)\left[y(k)-C \hat{x}(k  | k-1)\right] ,\\
P(k)&=P(k  | k-1)-K(k) C P(k  | k-1),
\end{align*}
where
\begin{align*}
&\hat{x}(k  | k-1)=A \hat{x}(k-1)+Bu(k), P(k  | k-1)=A P(k-1) A{'}+Q ,\\	
&K(k)=P(k  | k-1) C{'}\left(C P(k  | k-1) C{'}+R\right)^{-1},
\end{align*}
with initial condition $\hat{x}(0  |-1)=0,\ P(0  |-1)=\Sigma $.
It is well-known that for observable system, the estimation error covariance matrices $P(k)$ and the gain $K(k)$ will converge to
\begin{align*}
P \triangleq \lim _{k \rightarrow \infty} P(k),\ P_{+}=A P A{'}+Q ,\ K \triangleq P_{+} C{'}\left(C P_{+} C{'}+R\right)^{-1}.
\end{align*}
Since typically the control system will be running for an extended period of time, we focus on the case where the Kalman filter is in steady state, and thus the Kalman filter reduces to the following fixed-gain linear estimator:
\begin{equation}\label{eq:fix_gain_kalman}
\hat{x}(k+1)=(I-K C) \left(A \hat{x}(k)+Bu(k)\right)+K y(k+1) .
\end{equation}
Before introducing our work, we first recall some results in the previous work that decomposes the fix gain Kalman filter to local estimates and recovers it securely by an optimization problem.

\subsection{Preliminary Results}\label{sec:preli}
We introduce some preliminaries in this subsection which are fundamental to main results in this paper.
The following assumption is introduced to prevent system degradation.
\begin{assumption}\label{as:distinct_eigvalue}
	The matrix $A$ is invertible; $A-K C A$ has $n$ distinct eigenvalues. Moreover, $A-K C A$ and $A$ do not share any eigenvalue.
\end{assumption}
\begin{remark}
	Since the invertibility of A implies that $(A, CA)$
	is also observable, we can freely assign the poles of $A-KCA$ by choosing a proper gain $K$. Hence, $A-KCA$ can satisfy the condition in Assumption \ref{as:distinct_eigvalue} with a small estimation performance loss.
\end{remark}
Since $A-K C A$ has distinct eigenvalues, it can be diagonalized as:
\begin{equation}\label{eq:VLambda}
A-K C A=V \Pi V^{-1}.
\end{equation}
Define the eigenvalues of $A-KCA$ as $\pi_{1},\cdots,\pi_{n}$.
Consider local estimation $\zeta_{i}(k)$ which is the system response of sensor $i$. The local estimator satisfies the following dynamic:
\begin{equation}\label{eq:def_zeta}
\zeta_{i}(k+1)=\Pi \zeta_{i}(k)+\mathbf{1}_{n} y_{i}(k+1)+\left(G_i-\mathbf{1}_n C_i\right)Bu(k) ,
\end{equation}
where $C_i$ is $i$-th row of matrix $C$, and $G_i$ is defined as
\begin{equation}\label{eq:def_Gi}
G_{i} \triangleq\left[\begin{array}{c}
C_{i} A\left(A-\pi_{1} I\right)^{-1} \\
\vdots \\
C_{i} A\left(A-\pi_{n} I\right)^{-1}
\end{array}\right].
\end{equation}
The following lemma shows the relationship between $\zeta_i(k)$ and $G_ix(k)$.
%the difference between them converge to a stationary Gaussian process in the absence of attack.
\begin{lemma}\label{lm:epsilon}
	$\zeta_i(k)$ is stable estimation of $G_ix(k)$. Define their difference as $\epsilon_i(k)\triangleq\zeta_{i}(k)-G_ix(k)$, then $\epsilon_i(k)$ satisfies the following dynamics:
	\begin{align}
	\epsilon_{i}(k+1)= \Pi \epsilon_{i}(k)-\left(G_{i}-\mathbf{1}_{n} C_{i}\right) w(k) 
	+\mathbf{1}_{n} v_{i}(k+1)&+\mathbf{1}_{n} a_{i}(k+1) . \label{eq:epsilon}
	\end{align}
\end{lemma}
\begin{proof}
	According to the definition of $\zeta_{i}(k+1)$, one obatins
	\begin{align*}
	\epsilon_{i}(k+1)=&\Pi\zeta_{i}(k)+\mathbf{1}_n\left[C_i\left(Ax(k)+Bu(k)+w(k)\right)+v_i(k+1)+a_i(k+1)\right]\\
	&-\left(G_{i}-\mathbf{1}_{n} C_{i}\right)Bu(k)-G_i\left(Ax(k)+Bu(k)+w(k)\right)\\
	=&\Pi\zeta_{i}(k)-\left(G_iA-\mathbf{1}_nC_iA\right)x(k)- \left(G_{i}-\mathbf{1}_{n} C_{i}\right) w(k) 
	+\mathbf{1}_{n}\left( v_{i}(k+1)+ a_{i}(k+1) \right)
	\end{align*}
	Since it has been proved in \cite{liuxinghua-TAC2020} Corollary 1 that $G_iA-\mathbf{1}_nC_iA=\Pi G_i$, one can verify that equation \eqref{eq:epsilon} holds. 
\end{proof}
Define $\tilde{Q} \in \Cb^{m n \times m n}$ as the covariance of noise term $\left(G_{i}-\mathbf{1}_{n} C_{i}\right) w(k) -\mathbf{1}_{n} v_{i}(k+1)$ for all $i$, i.e.,
\begin{align}
\tilde{Q} \triangleq
\begin{bmatrix}
G_1-\mathbf{1}_{n} C_1 \\
\vdots \\
G_m-\mathbf{1}_{n} C_m
\end{bmatrix}
Q\begin{bmatrix}
G_1-\mathbf{1}_{n} C_1 \\
\vdots \\
G_m-\mathbf{1}_{n} C_m
\end{bmatrix}^{'}
+ R\otimes \mathbf{1}_{n\times n},
\end{align}
where $\otimes$ is the Kronecker product.
Define $\tilde{\Pi}\in \Cb^{m n \times m n}$ as
$$
\tilde{\Pi} \triangleq\left[\begin{array}{ccc}
\Pi & & \\
& \ddots & \\
& & \Pi
\end{array}\right].
$$
%\textbf{\footnote{$\epsilon(k)$ and $\epsilon_i(k)$ is defined in Lemma \ref{lm:epsilon}}}
The stable covariance of $\epsilon(k)\triangleq\left[\epsilon_1(k){'},\cdots,\epsilon_m(k){'}\right]{'}$ is the solution $\tilde{W}$ of the following Lyapunov equation:
$$
\tilde{W}=\tilde{\Pi} \tilde{W} \tilde{\Pi}{'}+\tilde{Q}.
$$ 
The matrix $\tilde{W}$ is well-defined since $\Pi$ is strictly stable. As a result, the secure estimation can be recovered by the solution of the following optimization problem where $\zeta(k)\triangleq\left[\zeta_1(k){'},\cdots,\zeta_m(k){'}\right]{'} $ and $G\triangleq\left[G{'}_1,\cdots,G{'}_m\right]{'} $.
\begin{subequations}\label{pb:old_lasso}
	\begin{align}
	\underset{\check{x}(k), \mu(k), \nu(k)}{\operatorname{minimize}}&\quad \frac{1}{2} \mu(k){'} \tilde{W}^{-1} \mu(k) + \gamma \|\nu(k)\|_1 \label{min:old_lasso} \\
	\text{subject to }&\quad
	\zeta(k)=
	G \check{x}(k)+\mu(k)+\nu(k). \label{eq:old_lasso}
	\end{align}
\end{subequations}
The parameter $\gamma$ is a non-negative constant chosen by the system operator. The following theorem from Liu et al.\cite{liuxinghua-TAC2020} proves that the solution $\check{x}(k)$ to problem (\ref{pb:old_lasso}) is a secure estimation under specific condition.
\begin{theorem}[Liu et al.\cite{liuxinghua-TAC2020}]\label{th:TAC}
	In the presence of $(p, m)$-sparse attack, the state estimation $\check{x}(k)$ is secure if the following inequality holds for all $x \neq \mathbf{0}$, $x\in\Rb^n$:
	\begin{equation}\label{eq:cond}
	\sum_{i \in \mathcal{I}}\left\|G_{i} x\right\|_{1}<\sum_{i \in \mathcal{I}^{c}}\left\|G_{i} x\right\|_{1}, \quad \forall\ \Ic\subset \Jc, |\mathcal{I}|\leq p .
	\end{equation}
\end{theorem}
Even though Theorem \ref{th:TAC} establishes the sufficient condition of the estimation to be secure, this condition can be improved in the following two aspects.
\begin{enumerate}[left=0pt]
	\item[(1)] Validating condition \eqref{eq:cond} is NP-hard. The computational complexity can be significantly reduced by introducing the non-derogatory assumption and further analysis on matrix $G_i$.
	\item[(2)] Condition \eqref{eq:cond} does not achieve the fundamental limit. It is more restrictive that $2p$-sparse observability and has a gap from $2p$-sparse detectablity.
\end{enumerate}

In the following section, we proposed a secure estimation scheme that improves the aforementioned two points.
Under assumption on unstable eigenvalues of $A$, the sufficient condition of the estimation to be secure is proved to be $2p$-sparse detectable, which is easily validated and achieves the fundamental limit of secure dynamic estimation.

\section{Secure Estimation with Sparse Detectability}\label{sec:main_result}
In this section, under the assumption that all the unstable eigenvalues of $A$ have geometric multiplicity $1$, we design a state estimator that is secure in the presence of $(p,m)$-sparse attack as long as the system $(A,C)$ is $2p$-sparse detectable. Moreover, we prove that if the system is not $2p$-sparse detectable, there exists an attack strategy under which no estimator can be secure.
%In this section, under the assumption on $A$, we first prove the relationship between span of row vectors of $G_i$ defined in \eqref{eq:def_Gi} and observable space of $(A,C_i)$. Leveraging the structure of $G_i$ and careful design of the optimization problem, we prove that the estimation is secure to $p$ compromised sensors if the system is $2p$-sparse detectable.
We first introduce the following assumption on unstable\footnote{Unstable eigenvalues are those eigvalues that satisfy $|\lambda_i| \geq1$.} eigenvalues of $A$.

\begin{assumption}\label{as:geo_unstable}
	All the unstable eigenvalues of $A$ have geometric multiplicity $1$. 
\end{assumption}
Since we can perform invertible linear transformation $T$ on state $x$ and study the following system instead:
\begin{align*}
\bar{x}(k)&= \bar{A} \bar{x}(k) + TB\bar{x}(k)+Tw(k),  \\
y(k)&=CT^{-1}\bar{x}(k)+v(k)+a(k) ,
\end{align*}
where $\bar{A}\triangleq TAT^{-1}$ is similar to $A$ and $\bar{x}=Tx$, we can assume that $A$ is in the following Jordan form without loss of generality:
\begin{align*}
&A=
\begin{pmatrix}
\begin{array}{cc}
A_1 & \mathbf{0} \\
\mathbf{0} & A_2			
\end{array}
\end{pmatrix}, \
%	A_\Uc=\begin{pmatrix}
%	J_{1} & \mathbf{0} & \cdots & \mathbf{0} \\
%	\mathbf{0} & J_{2} & \cdots & \mathbf{0} \\
%	\vdots & \vdots & \ddots & \vdots \\
%	\mathbf{0} & \mathbf{0} & \cdots & J_{l} 
%	\end{pmatrix},	\\
%	&
%	\text{where } J_k=
%	\begin{pmatrix}
%	\lambda_{k} & 1 & 0 & \cdots & {0} \\
%	{0} & \lambda_{k} & 1 &  \cdots & {0} \\
%	\vdots & \vdots &  \vdots & \ddots & \vdots \\
%	{0} & {0} & {0} & \cdots & \lambda_{k} 
%	\end{pmatrix} ,
%\in \Cb^{n_k \times n_k} , \sum_{k=1}^{l}n_k =|\Uc| .
\end{align*}
where block $A_1$ is composed of the Jordan blocks with unstable eigenvalues and $A_2$ is composed of the Jordan blocks with stable eigenvalues.

Denote the number of unstable eigenvalues of $A$ (counted with repetition) as $n_u$ and number of stable eigenvalues as $n_s$.
In order to analyze stable and unstable states separately with simple notation, we denote the index set of unstable entries of state as $\Uc\triangleq\{1,2,\cdots,n_u\}$ and index set of stable entries as $\Sc\triangleq\{n_u+1,\cdots,n\}$. 
Furthermore, a matrix $X$ can be divided vertically to two sub-matrices:
\begin{equation*}
X=\left[\begin{array}{c} X^\Uc \ | \ \ X^\Sc\end{array}\right] ,
\end{equation*}
where $X^\Uc$ is the matrix composed of first $n_u$ columns of matrix $X$ and $X^\Sc$ is composed of last $n_s$ columns of $X$. 
%For example, according to Assumption \ref{as:geo_unstable}, $A^\Uc=\begin{bmatrix} A_1\\ \mathbf{0} \end{bmatrix}, A^\Sc=\begin{bmatrix} \mathbf{0} \\ A_2 \end{bmatrix}$.
Define the observable matrix of system $(A,C_i)$ as 
\begin{equation}\label{eq:def_O}
O_{i} \triangleq\left[\begin{array}{c|c|c|c}
C_{i}{'} &
\left(C_{i} A\right){'} &
\cdots &
\left(C_{i} A^{n-1}\right){'}
\end{array}\right]{'}.
\end{equation}
Therefore, $\rs(O^\Uc_{i})$ is the observable subspace of sensor $i$ corresponding to unstable states.

\subsection{Canonical form of $G_i$}\label{subsec:transform}
We prove in this subsection that the row span of $G^\Uc_i$ coincides with $\rs(O^\Uc_{i})$, which implies that the matrix $G^\Uc_i$ has a canonical form under row operations. 
Before continuing on, we need the following notation of state-sensor observability. 
%Define $\Sc\triangleq \{1,2,\cdots,n\}$ as the index set of all states and $\Sc_i\subseteq\Sc$ as the index set of states that sensor $i$ can observe, i.e.,
%\begin{equation}
%	\Sc_i\triangleq \{j\in\Sc\ |\ O_i{'} e_j\neq \mathbf{0} \},
%\end{equation}
Define $\Ec_j$ as the index set of sensors that can observe state $j$, i.e.
\begin{equation}\label{eq:def_Ec}
\Ec_j\triangleq \{i\in\Jc\ |\ O_i{'} e_j\neq \mathbf{0} \},
\end{equation}
where $\Jc\triangleq \{1,2,\cdots,m\}$ is the index set of all sensors and $e_j$ is the $n$-dimensional canonical basis vector with 1 on the $j$-th entry and 0 on the other entries.
We have the following theorem characterizing the structure of $G_i$. 
\begin{theorem}\label{th:span}
	Assume system matrix $A$ satisfies Assumption \ref{as:geo_unstable}, then the following equation holds:
	\begin{align}\label{eq:span}
	\rs(G_i^\Uc)=\rs(O_i^\Uc)=\rs(H_i^\Uc) ,
	\end{align}
	where $H^\Uc_i$ is the following $n\times n_u$ matrix
	\begin{equation*}
	H_i^\Uc\triangleq \begin{bmatrix}
	\Ib_{i\in\Ec_1}\hspace{-25pt} & & \\
	& \ddots&   \\
	& & \hspace{-20pt}\Ib_{i\in\Ec_{n_u}}\\
	\hline \\
	&\hspace{10pt}\mathbf{0}_{(n-n_u)\times n_u} &\\
	& &
	\end{bmatrix} ,
	\end{equation*}
	and $\mathbb{I}_\Es$ is the indicator function that takes the value 1 when $\Es$ is true and value 0 when $\Es$ is not.
	Therefore, there exists an invertible $n\times n$ matrix $P_i$ such that $P_iG_i=H_i$ and $H_i$ is in the following form:
	\begin{equation*}
	H_i=P_iG_i=\left[\begin{array}{c} H_i^\Uc \ | \ \ P_i G^\Sc_i\end{array}\right] .
	\end{equation*}
\end{theorem}

%Theorem \ref{th:span} directly follows Theorem 2 in \cite{} and is omitted because of space limit.
Proof of Theorem \ref{th:span} is provided in Appendix \ref{ap:span}.
After transformation $P_i$, matrix $G^\Uc_i$ is transformed into canonical form $H^\Uc_{i}$ whose rows are either canonical basis vectors or zero vectors. 
The non-zero entries of $H_i$ records the state observability of sensor $i$. Therefore, the sparse observability/detectability index can be directly obtained from $H^\Uc_i$.
\begin{corollary}\label{co:sparse_obs}
	The sparse observability index of system $(A,C)$ is $\min_{j\in\{1,2,\cdots,n\}} \left|\Ec_{j}\right| - 1 $.
	The sparse detectability index of system $(A,C)$ is $\min_{j\in\Uc} \left|\Ec_{j}\right| - 1 $ if $\Uc\neq\varnothing$ and is $m-1$ if $\Uc=\varnothing$.
\end{corollary} 
%\begin{remark}
%	Since we focus on observable system, then for each $j\in\{1,2,\cdots,n\}$, $\Ec_j\neq\varnothing$. Thus, the sparse observability index and sparse detectability index are non-negative integers.
%\end{remark}
\begin{proof}
	Define $s\triangleq \min_{j\in\{1,2,\cdots,n\}} \left|\Ec_{j}\right| - 1.$
	For arbitrary $\overline{s}$ that satisfy $\overline{s}\geq s+1$, there exists a state index $j^*$ and a sensor index set $\Ic^*$ with $|\Ic^*|=\overline{s}$ such that $\Ec_{j^*} \cap \left(\Jc\setminus\Ic^*\right)=\varnothing$.
	As a result, state $j^*$ cannot be observed by any sensor in $\Jc\setminus \Ic^*$, i.e.,
	\begin{equation*}
	e_{j^*}\notin \rs(O_i),\ \forall i\in\Jc\setminus \Ic^*,
	\end{equation*}
	and thus system $(A,C_{\Jc\setminus\Ic^*})$ is not observable.
	For arbitrary $\underline{s}$ that satisfies $\underline{s}\leq s$, arbitrary $j$ and arbitrary $\Ic$ with $|\Ic|=\underline{s}$, one obtains $\Ec_{j^*} \cap \left(\Jc\setminus\Ic^*\right)\neq\varnothing$, which means for all $j$, there exists $i^*\in\Jc\setminus \Ic$ such that: $e_{j}\in \rs(O_{i^*})$. Therefore, system $(A,C_{\Jc\setminus\Ic})$ is observable. According to Definition \ref{df:sparse_obs}, the system is $s$-sparse observable. The detectability index is obtained in the same way by considering unstable subsystem when $\Uc\neq \varnothing$. When $\Uc=\varnothing$(i.e., $A$ is stable), system is always detectable as long as $\Jc\neq\varnothing$ and thus the sparse detectability is $m-1$ according to definition \ref{df:sparse_obs}. 
\end{proof}

In conclusion, under Assumption \ref{as:geo_unstable}, the matrix $G^\Uc_i$ has a canonical form which $H^\Uc_i$ is determined by state-sensor observability. Leveraging upon the canonical form $H^\Uc_i$, we will propose an estimation scheme that is secure in the presence of $(p,m)$-sparse attack as long as the system is $2p$-sparse detectable.

\subsection{Secure Estimation Design}
Recalling the transformation $P_i$ introduced in Theorem \ref{th:span}, define $\tilde{P} \triangleq \text{diag}\left(P_1,\cdots,P_m\right)$, $\tilde{M}\triangleq\tilde{P}\tilde{W}\tilde{P}{'}$ and
\begin{align}\label{eq:def_YH}
{Y} (k)\triangleq
\begin{bmatrix}
P_1\zeta_{1}(k) \\
\vdots \\
P_m\zeta_{m}(k)
\end{bmatrix}\in\Cb^{mn\times 1}, \
H\triangleq\begin{bmatrix}
H_{1} \\
\vdots \\
H_{m}
\end{bmatrix}\in\Cb^{mn\times n} .	
\end{align}
Define the following matrix
\begin{align*}
&N\triangleq
I_{m} \otimes 
\begin{bmatrix}
\mathbf{0}_{n_s\times n_u} & I_{n_s}
\end{bmatrix}
\in \Rb^{mn_s\times mn }.
\end{align*}
Consider the following least square problem.
\begin{subequations}\label{pb:least_square}
	\begin{align}
	\underset{{\tilde{x}_\ls}(k), \varphi(k)}{\text{minimize}}&\quad \frac{1}{2} 
	\begin{bmatrix}
	\varphi(k) \\
	N H \tilde{x}_\ls(k)
	\end{bmatrix}^{'} \Wc
	\begin{bmatrix}
	\varphi(k) \\
	N H \tilde{x}_\ls(k)
	\end{bmatrix}  \\
	\text {subject to}&\quad
	{Y} (k)= H \tilde{x}_\ls(k)+\varphi(k).  
	\end{align}
\end{subequations}
where 
\begin{align}\label{eq:def_W}
\Wc\triangleq \begin{bmatrix}
\tilde{M}^{-1}+ N{'}N & N{'} \\
N &  I
\end{bmatrix}.
\end{align}
Notice that $\Wc$ is positive definite since $\tilde{M}^{-1}\succ 0$.
Define\footnote{$\diag(V^{-1}K_i)$ is a $n\times n$ diagonal matrix whose diagonal with the $j$-th diagonal entry equals to $j$-th element of vector $V^{-1}K_i$.} 
$F_i\triangleq V\diag(V^{-1}K_i),\ F=\begin{bmatrix} F_1&\cdots & F_m \end{bmatrix}$,
where $V$ is defined in \eqref{eq:VLambda}.
Recall that $\epsilon(k)\triangleq\left[\epsilon_1(k){'},\cdots,\epsilon_m(k){'}\right]{'}$ and $\epsilon_i(k)=\zeta_i(k)-G_ix(k)$ from Lemma \ref{lm:epsilon} and fix gain Kalman estimation $\hat{x}(k)$ from \eqref{eq:fix_gain_kalman}.
\begin{lemma}\label{lm:least_square}
	In the absence of attack,  the solution to least square problem \eqref{pb:least_square} coincides with the Kalman estimation and satisfies the following:
	\begin{equation*}
	\tilde{x}_\ls(k)=\hat{x}(k),\ \varphi(k)=(I-GF)\epsilon(k).
	\end{equation*}
\end{lemma}
\begin{proof}
	Consider the following least square problem
	\begin{align}\label{pb:least_square_orign}
	\underset{\tilde{x}_\ls (k)}{\operatorname{minimize}}\quad \frac{1}{2} \left({Y}(k)-H \tilde{x}_\ls(k)\right){'} \tilde{M}^{-1} \left({Y}(k)-H \tilde{x}_\ls(k)\right) . 
	\end{align}
	Based on Theorem 2 in \cite{liuxinghua-TAC2020}, the solution to problem \eqref{pb:least_square_orign} is equivalent to Kalman estimation. It is sufficient to prove that problems \eqref{pb:least_square} and \eqref{pb:least_square_orign} are equivalent.
	Define 
	$$
	\Mc\triangleq
	\begin{bmatrix}
	I_{mn} & \mathbf{0}_{mn\times mn_s} \\ 
	N		
	& I_{mn_s} 
	\end{bmatrix}
	\in \Rb^{m(n+n_s)\times m(n+n_s) }.
	$$
	Consider the objective function of problem \eqref{pb:least_square_orign} added by a constant term\footnote{$Y(k)$ is fixed for each $k$ in the optimization problem and thus is treated as a constant. For legibility, the time index $(k)$ is omitted.}:
	\begin{align}
	&\frac{1}{2} ({Y} - H\tilde{x}_{\ls} ){'} \tilde{M}^{-1} ({Y} - H\tilde{x}_{\ls} ) +\frac{1}{2}  Y'N'N Y =\frac{1}{2}\begin{bmatrix}
	{Y} - H\tilde{x}_{\ls} \\ N Y
	\end{bmatrix}^{'}
	\begin{bmatrix}
	\tilde{M}^{-1} & \mathbf{0} \\
	\mathbf{0} &  I
	\end{bmatrix}
	\begin{bmatrix}
	{Y} - H\tilde{x}_{\ls} \\ N Y
	\end{bmatrix}. \label{eq:expand_mn_to_mn+ms}
	\end{align}
	Notice that 
	\begin{equation*}
	\begin{bmatrix}
	{Y} - H\tilde{x}_{\ls} \\
	N Y
	\end{bmatrix}=\Mc
	\begin{bmatrix}
	{Y} - H\tilde{x}_{\ls} \\
	N H \tilde{x}_{\ls}
	\end{bmatrix},
	\end{equation*}
	and (\ref{eq:expand_mn_to_mn+ms}) can be written as 
	\begin{align}\label{eq:obj_function}
	&\frac{1}{2}
	\left(\Mc
	\begin{bmatrix}
	{Y} - H\tilde{x}_{\ls} \\
	N H \tilde{x}_{\ls}
	\end{bmatrix}
	\right)^{'}
	\begin{bmatrix}
	\tilde{M}^{-1} & \mathbf{0} \\
	\mathbf{0} &  I
	\end{bmatrix}
	\left(\Mc
	\begin{bmatrix}
	{Y} - H\tilde{x}_{\ls} \\
	N H \tilde{x}_{\ls}
	\end{bmatrix}
	\right) 
	=\frac{1}{2}
	\begin{bmatrix}
	{Y} - H\tilde{x}_{\ls} \\
	N H \tilde{x}_{\ls}
	\end{bmatrix}^{'}
	\Wc
	\begin{bmatrix}
	{Y} - H\tilde{x}_{\ls} \\
	N H \tilde{x}_{\ls}
	\end{bmatrix} . 
	\end{align}
	Substituting $\varphi$ in \eqref{pb:least_square} with ${Y}-H\tilde{x}_{\ls}$ leads to \eqref{eq:obj_function}. Thus, optimizing objective function \eqref{pb:least_square_orign} is equivalent to optimizing \eqref{eq:obj_function}, and the latter is equivalent to problem \eqref{pb:least_square}.   
\end{proof}

Based on least square problem \eqref{pb:least_square}, we present the following optimization problem whose solution $\tilde{x}(k)$ is our proposed secure estimation. The constant $\gamma$ is a non-negative adjustable parameter.
\begin{subequations}\label{pb:resilient_LASSO}
	\begin{align}
	\underset{{\tilde{x}}(k), \mu(k),\nu(k)}{\text{minimize}}&\quad \frac{1}{2} 
	\begin{bmatrix}
	\mu(k) \\
	N H \tilde{x}(k)
	\end{bmatrix}^{'} \Wc
	\begin{bmatrix}
	\mu(k) \\
	N H \tilde{x}(k)
	\end{bmatrix} + \gamma\left\|\nu(k)\right\|_1  \\
	\text { subject to }&\quad
	{Y} (k)= H \tilde{x}(k)+\mu(k)+\nu(k) .  
	\end{align}
\end{subequations}

The following theorem characterizes the performance of our proposed estimator when the attacker is absent. 
The proof is provided in Appendix \ref{ap:main}.
\begin{theorem}\label{th:no_attack}
	In the absence of attack, if the parameter $\gamma$ in problem (\ref{pb:resilient_LASSO}) satisfies
	\begin{align}\label{eq:kalman_cond}
	\left\|\Wc \begin{bmatrix}
	\left(I-GF\right)\epsilon(k) \\
	N H \hat{x}(k)
	\end{bmatrix}\right\|_\ift\leq\gamma, 
	\end{align}
	then our proposed estimation $\tilde{x}(k)$ is equivalent to the estimation of fixed gain Kalman filter defined in (\ref{eq:fix_gain_kalman}), i.e.,
	\begin{equation}\label{eq:eq_to_kalman}
	\tilde{x}(k)=\hat{x}(k).
	\end{equation}
\end{theorem}
Noticing that $\epsilon(k)$ converges to a stationary Gaussian process, and $\hat{x}(k)$ is a Gaussian random variable, the probability that \eqref{eq:kalman_cond} holds is determined only by system parameter $A,B,C,Q,R,\gamma$ given input $u(k)$, and can be explicitly calculated given these parameters.
By tuning design parameter $\gamma$, the probability of recovering the Kalman estimation can be adjusted.

In order to quantify the estimation difference between the attack is absent and present, we consider the following local estimation and Kalman estimation without attack:
\begin{align}
\zeta^o_i (k+1) &= \Pi \zeta^o_i (k) + \mathbf{1}_n z_i (k+1) + (G_i-\mathbf{1}_nC_i)Bu(k), \label{eq:def_zetare} \\
\hat{x}^o (k+1)& = (I-K C) \left(A \hat{x}^o(k)+Bu(k)\right) + K z (k+1),\label{eq:def_xhatre}
\end{align}
where $z(k)=Cx(k)+v(k)$ is the original (unmanipulated) measurement.
Define $\epsilon^o_{i}(k)$ correspondingly as $\epsilon^o_{i}(k)\triangleq \zeta^o_i(k)-G_ix(k)$.
The following theorem quantifies the estimation error introduced by the attack.

\begin{theorem}\label{th:main}
	In presence of arbitrary $(p,m)$-sparse attack, if the system $(A,C)$ is $2p$-sparse detectable, then 
	the estimation difference between $\tilde{x}(k)$ solved from (\ref{pb:resilient_LASSO}) and oracle Kalman estimation $\hat{x}^o(k)$ satisfies
	\begin{equation}\label{eq:diff_upper_bound}
	\left|[\tilde{x}(k)]_j-[\hat{x}^o(k)]_j\right|\leq
	\begin{cases}
	\underset{i_1,i_2\in \Ec_j}{\max} \left| \left[P_{i_1} \zeta^o_{i_1}(k)\right]_j- \left[P_{i_2} \zeta^o_{i_2}(k)\right]_j\right| + \left(\gamma+\gamma^o(k) \right) \left\|\Fc\right\|_\ift ,\ j\in\Uc, \\
	\gamma \cdot \left\|\Fc\right\|_\ift+\left| [\hat{x}^o(k)]_j \right|,\ j\in\Sc,
	\end{cases}
	%	\|\tilde{x}(k)-\hat{x}^o(k)\|_\ift\leq \Gamma(k)+\left(\gamma+\gamma^o(k) \right) \left\|\Fc\right\|_\ift,
	\end{equation}
	where
	\begin{align*}
	\gamma^o(k)&\triangleq \left\|\Wc \begin{bmatrix}
	\left(I-GF\right)\epsilon^o(k) \\
	N H \hat{x}^o(k)
	\end{bmatrix}\right\|_\ift,\\
	\Fc	&\triangleq
	\left(
	\begin{bmatrix}
	I_{mn} & \mathbf{0} \\
	\mathbf{0}  &  \Lc H{'} N{'}
	\end{bmatrix}
	\Wc
	\begin{bmatrix}
	I_{mn} & \mathbf{0} \\
	\mathbf{0}  &  N H \Lc{'}
	\end{bmatrix}
	\right)^{-1}
	\begin{bmatrix}
	I_{mn} & \mathbf{0} \\ \mathbf{0} &\Lc H{'}
	\end{bmatrix},\\
	\Lc&\triangleq 
	\begin{bmatrix}
	\mathbf{0}_{n_s\times n_u} & I_{n_s}
	\end{bmatrix},
	\end{align*}
	with $\Ec_j$ defined in \eqref{eq:def_Ec} and $[\cdot]_j$ being the $j$-th element of a vector.
	Since the oracle Kalman estimation $\hat{x}^o(k)$ is a stable estimation of system state $x(k)$, and the upper bounds have bounded variance for all $k\in\Zb^+$, our proposed estimation $\tilde{x}(k)$ is secure. 		
\end{theorem}

Under Assumption \ref{as:geo_unstable}, if the system is $2p$-sparse detectable, our proposed estimator is secure. The maximum estimation difference from oracle Kalman filter is shown in \eqref{eq:diff_upper_bound}.
Theorem \ref{th:main} indicates smaller $\gamma$ leads to lower estimation difference upper bound in the presence of attack. However, based on Theorem \ref{th:no_attack}, smaller $\gamma$ decreases probability of recovering the optimal Kalman estimation in the absence of attack. The choice of $\gamma$ represents the trade-off between the performance in normal operation and the performance under attack. 

Moreover, since sparse detectability index only requires simple computation according to Corollary \ref{co:sparse_obs}, our work reduces the complexity of evaluating system vulnerability significantly under the assumption of geometric multiplicity.
For general $A$ that has eigenvalues with geometric multiplicity larger than 1 ($A$ is derogatory), computing sparse observability is an NP-hard problem \cite{mao2021computational}, and there is no computational efficient solution unless $\rm P=NP$. Besides the computation complexity of off-line designing, for algorithm online operation, the computing of estimation involves solving a convex optimization problem which can be done efficiently.

The following theorem proves that $2p$-sparse detectability is necessary for the existence of a secure estimation, which coincides with the sufficient condition of our estimator to be secure.
% which 
%indicates our proposed estimation scheme provides a secure estimation whenever it is possible in the presence of $(p,m)$-sparse attack.
\begin{theorem}\label{th:funda_lim} %[\hspace{-0.001pt}\cite{yorie_TAC2018} Theorem 1]
	If the system is not $2p$-sparse detectable, there exists a $(p,m)$-sparse attack strategy under which no estimator is secure.
\end{theorem}

\begin{proof}
	This proof is based on \cite{yorie_TAC2018}\cite{nakahira2018attackresilient}. Since we focus on a different formulation, we reorganize it here for paper self-consistency.
	If the system is not $2p$-sparse detectable, then there exists a set $\Ac$ with $\Ac\subset\Jc,|\Ac|=2p$ and an eigenvector $\xi$ of $A$ that corresponds to a unstable eigenvalue $\lambda$ ($|\lambda|\geq 1$) such that either one of the following two statements is true:
	\begin{align*}
	&\text{(1) } A\xi=\lambda \xi, C_{\Jc\setminus\Ac}\xi=\mathbf{0}, \text{if $\xi$ is a real-value vector},\\
	&\text{(2) } A\xi=\lambda \xi,A\bar{\xi}=\bar{\lambda} \bar{\xi}, C_{\Jc\setminus\Ac}\xi=\mathbf{0},C_{\Jc\setminus\Ac}\bar{\xi}=\mathbf{0}, \text{if $\xi$ is a complex-value vector},
	\end{align*} 
	where $\bar{\xi},\bar{\lambda}$ represent the conjugate of $\xi,\lambda$.
%	As a result, there exists a set $\Ac$ with $\Ac\subset\Jc,|\Ac|=2p$, such that the linear transformation defined by $\Oc_t$ has a non-trivial kernel, where  
%	$$\Oc_{t}\triangleq\left[\begin{array}{c|c|c|c}
%	C_{\Jc\setminus\Ac}{'} &
%	\left(C_{\Jc\setminus\Ac} A\right){'} &
%	\cdots &
%	\left(C_{\Jc\setminus\Ac} A^{t-1}\right){'}
%	\end{array}\right]{'}.$$
%	In other words, either one of the following two statements is true.
%	\begin{align*}
%	&\text{(1) if $\xi$ is a real vector: }\Oc_t\xi=\mathbf{0} ,  \forall t\geq 0. \\
%	&\text{(2) if $\xi$ is a complex vector:  }\Oc_t\left(\xi+\bar{\xi}\right)=\mathbf{0} ,  \forall t\geq 0.
%	\end{align*} 
	Based on this result, we intend to prove that the following proposition is true.
	\begin{proposition}\label{prop}
		There exists zero mean Gaussian disturbances $\{w^{(1)}(k)\}_{k=0}^{\ift}$, $\{v^{(1)}(k)\}_{k=0}^{\ift}$, $\{w^{(2)}(k)\}_{k=0}^{\ift}$, $\{v^{(2)}(k)\}_{k=0}^{\ift}$, two $(p,m)$-sparse attack sequences $\{a^{(1)}(k)\}_{k=0}^{\ift},\{a^{(2)}(k)\}_{k=0}^{\ift}$ and a pair of initial states $x^{(1)}(0),x^{(2)}(0)$ such that the two system trajectories $\{x^{(1)},w,y^{(1)},v,a^{(1)}\},\{x^{(2)},w,y^{(2)},v,a^{(2)}\}$ satisfy:
		\begin{itemize}
			\item Two system trajectories both follow dynamics in \eqref{eq:system}-\eqref{eq:y_i_def}.
			\item $y^{(1)}(k)=y^{(2)}(k),\ \forall k\geq 0$.
			\item $\|x^{(1)}(k)-x^{(2)}(k)\|_2\ra\ift$, as $k\ra\ift$.
		\end{itemize}
	\end{proposition}
	We construct two trajectories that proves Proposition \ref{prop}. 
	Divide $\Ac$ into $\Ac=\Ac_1\cup\Ac_2$ such that $\Ac_1\cap\Ac_2=\varnothing$ and $|\Ac_1|=|\Ac_2|=p$.
	Define the following two trajectories if $\xi$ is real-value vector:
	$$
	\begin{array}{llll}
	\textbf{System 1:}&	x^{(1)}(0)=\mathbf{0} ,  & \textbf{System 2:} &x^{(2)}(0)=\xi ,\\
	& w^{(1)}(k)=\mathbf{0},\ v(k)=\mathbf{0}, & &  w^{(2)}(k)=\phi(k)\cdot\xi,\ v(k)=\mathbf{0},\\
	&	a^{(1)}(k)=\begin{cases}
	C_i x^{(2)}(k),i\in\Ac_1\\
	\mathbf{0},i\in\Jc\setminus\Ac_1
	\end{cases}, & &
	a^{(2)}(k)=\begin{cases}
    - C_i x^{(2)}(k),i\in\Ac_2\\
	\mathbf{0},i\in\Jc\setminus\Ac_2
	\end{cases}.
	\end{array}
	$$
	where $\phi(k)$ is a time-independent, standard Gaussian distributed random scalar.
	Noticing that $A\xi=\lambda \xi$, $C_{\Jc\setminus\Ac}\xi=\mathbf{0}$, one can verify that the output of these two systems are:
	\begin{align*}
	\begin{array}{llll}
		\textbf{System 1:}
		&	y^{(1)}(k)=\begin{cases}
			C_i x^{(2)}(k),i\in\Ac_1\\
			\mathbf{0},i\in\Ac_2\\
			\mathbf{0},i\in\Jc\setminus\Ac
		\end{cases}, & \textbf{System 2:} &
		y^{(2)}(k)=\begin{cases}
			C_i x^{(2)}(k),i\in\Ac_1\\
			\mathbf{0},i\in\Ac_2\\
			\mathbf{0},i\in\Jc\setminus\Ac
		\end{cases}.
	\end{array}
	\end{align*}
	The state sequence of system 2 satisfies
	$$
	x^{(2)}(k)=   \lambda^k\xi + \sum_{t=1}^{k} \lambda^{k-t}\xi \phi(t)
	$$
	while state sequence of system 1 is $x^{(1)}(k)=\mathbf{0}$.
	One concludes that $\Eb\|x^{(1)}(k)-x^{(2)}(k)\|^2_2\ra\ift$ as $k\ra\ift$ for both $|\lambda|>1$ and $|\lambda|=1$.

	If $\xi$ is complex, let $x^{(2)}(0)=\xi+\bar{\xi}$, $w^{(2)}(k)=\phi(k)\left(\xi+\bar{\xi}\right)$ and others hold the same with real-value case.
	One can verify that 
	$$
	x^{(2)}(k)=   \left(\lambda^k\xi+\bar{\lambda}^k\bar{\xi}\right) + \sum_{t=1}^{k} \left(\lambda^{k-t}\xi+\bar{\lambda}^{k-t}\bar{\xi}\right) \phi(t)
	$$
	and $x^{(1)}(k)=\mathbf{0}$.
	As a result, $\Eb\|x^{(1)}(k)-x^{(2)}(k)\|^2_2\ra\ift$ as $k\ra\ift$ for both $|\lambda|>1$ and $|\lambda|=1$.
	
	However, the system observable output $y^{(1)}(k)$ and $y^{(2)}(k)$ are always the same.
	As a result, Proposition \ref{prop} is proved and there exists no secure estimation under such attack. 
\end{proof}

In view of Theorem \ref{th:funda_lim}, our proposed estimator achieves the fundamental limit of secure estimation problem, i.e., provides a secure estimation whenever the system is possible to be securely estimated.
This result is stronger than other dynamic estimators in the literature which require $2p$-sparse observability \cite{FawziTAC2014,Pajic2017TAC,chang2018auto,Shoukry2017,Mishra2017TCNS,liuxinghua-TAC2020}.
The performance of our proposed estimator is corroborated by the numerical simulation in the next section.

\section{Illustrative Example}\label{sec:sim}
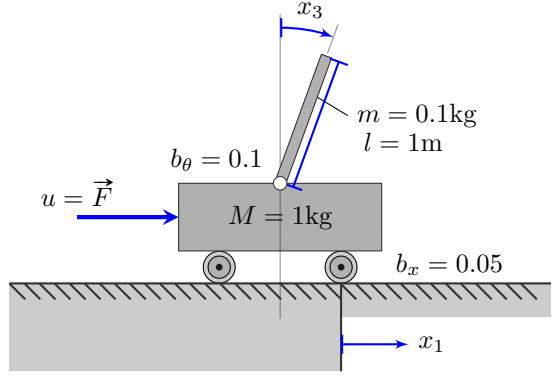
\begin{figure}[htpb]
	\centering
	%\tikzsetnextfilename{PIDBode}
\begin{tikzpicture}[> = latex',%
	scale = 0.9]
	\tikzset{%
		interface/.style={
			% The border decoration is a path replacing decorator. 
			% For the interface style we want to draw the original path.
			% The postaction option is therefore used to ensure that the
			% border decoration is drawn *after* the original path.
			postaction={draw, decorate, decoration={border, angle=-45,
					amplitude=0.3cm, segment length=2mm}}},
		helparrow/.style={>=latex', blue, thick},
		% helparrow/.style={>=latex', draw=blue, fill=blue, very thick},
		helpline/.style={thin, black!90, opacity=0.5},
		force/.style={>=stealth, draw=blue, fill=blue, ultra thick},
	}
	\def\ground{%
		\fill [black!20] (0, 0) rectangle (49mm, -13mm);
		\fill [black!20] (49mm, 0) rectangle (80mm, -5mm);
		\draw [thick, black!80, interface] (0, 0) -- (80mm, 0);
		\draw [thick, black!80] (49mm, 0) -- +(0, -13mm);
		\draw [|->, helparrow] (49mm, -9mm) -- ++(10mm, 0) node [right] {\color{black} $x_1$ };
	}
	
	\def\cart{
		\filldraw [%thick,%
		draw = black!80,%
		fill = black!40,%
		top color = black!30,%
		bottom color = black!30,%
		% pattern=horizontal lines gray,%
		] (0,0) rectangle (30mm, 10mm);
		\node at (15mm, 5mm) {$M=1$kg};
		\draw[->, force] (-15mm, 5mm) node [above] {$u=\ensuremath{\vv{F}}$} -- (0, 5mm);
		\node at (40mm, -2mm) {$b_x = 0.05$};
	}
	
	\def\pendulum{%
		\filldraw [%thick,%
		draw = black!80,%
		% fill = black!25,%
		left color=black!30,%
		bottom color=black!30,%
		% pattern=horizontal lines gray,%
		] (-0.8mm, 0) rectangle (0.8mm, 20mm);
		\draw [|-|, helparrow] (2mm, 0mm) -- ++(0mm, 20mm);
		\node at (15mm, 12mm) {$l=1$m};
		\node at (-10mm, 0mm) {$b_\theta=0.1$}; 
	}
	
	\def\joint{%
		\filldraw [%thick,%
		draw = black!80,%
		fill = white,%
		% pattern=horizontal lines gray,%
		] (0, 0) circle (1mm);
	}
	
	\def\wheel{%
%		\fill [thin,%
%		fill = black!70,%	path fading = south
%	] (0, 0) circle (1.75mm);
%		\begin{scope}
%			\clip (0, 0) circle (1.75mm);
%			\fill [fill=black!30,] (0, -1mm) circle (2mm);
%		\end{scope}
		\fill [fill=black!30,] (0, 0mm) circle (2mm);
		\fill [fill=black!90] (0, 0) circle (0.5mm);
		\draw [thin,%
		double = black!20,%
		double distance = 0.5mm] (0, 0) circle (2mm);
	}

	\begin{scope}
		\wheel
	\end{scope}
	\begin{scope}[xshift=18mm]
		\wheel
	\end{scope}
	\begin{scope}[xshift=-31mm, yshift=-2.4mm]
		\ground
	\end{scope}
	\begin{scope}[shift = {(-6mm, 2.4mm)}]
		\cart
	\end{scope}
	\begin{scope}[shift = {(9mm, 12.4mm)}]
		\draw [helpline] (0, -20mm) -- (0, 25mm);
		\draw [helpline] (70:20.5mm) -- (70:25mm);
		\draw [|->, helparrow] (0, 23mm) 
		arc [radius=23mm, start angle=90, delta angle=-20] ;
		\node at (80:26mm) {$x_3$};
%		\draw [|->, helparrow] (0, -9mm) 
%		arc [radius=9mm, start angle=-90, delta angle=195] ;
%		\node [above right] at (50:8.5mm) {$\theta$};
		\draw [thin] (70:14mm) -- (10mm, 10mm) 
		node [right] {$m=0.1$kg};
	\end{scope}
	\begin{scope}[shift = {(9mm, 12.4mm)}, rotate=-20]
		\pendulum
	\end{scope}
	\begin{scope}[shift = {(9mm, 12.4mm)}]
		\joint
	\end{scope}
\end{tikzpicture}

%\begin{tikzpicture}[> = latex',%
%	scale = 1,%
%	text = blue]
%	\begin{scope}
%		\wheel
%	\end{scope}
%	\begin{scope}[xshift=18mm]
%		\wheel
%	\end{scope}
%	\begin{scope}[xshift=-31mm, yshift=-2.4mm]
%		\ground
%	\end{scope}
%	\begin{scope}[shift = {(-6mm, 2.4mm)}]
%		\cart
%	\end{scope}
%	\begin{scope}[shift = {(9mm, 12.4mm)}]
%		\draw [helpline] (0, -20mm) -- (0, 25mm);
%		\draw [helpline] (110:20.5mm) -- (110:25mm);
%		\draw [|->, helparrow] (0, 23mm) 
%		arc [radius=23mm, start angle=90, delta angle=20] ;
%		\node at (97:25mm) {$\theta'$};
%		\draw [|->, helparrow] (0, -9mm) 
%		arc [radius=9mm, start angle=-90, delta angle=195] ;
%		\node [above right] at (50:8.5mm) {$\theta$};
%		\draw [thin] (110:14mm) -- (-10mm, 10mm) 
%		node [left] {$m$, $I$};
%	\end{scope}
%	\begin{scope}[shift = {(9mm, 12.4mm)}, rotate=20]
%		\pendulum
%	\end{scope}
%	\begin{scope}[shift = {(9mm, 12.4mm)}]
%		\joint
%	\end{scope}
%\end{tikzpicture}
	\caption{Illustration of the inverted pendulum.}\label{fig:invpen}
\end{figure}

We use an inverted pendulum for the numerical simulation\footnote{The corresponding code is posted on \texttt{https://github.com/zs-li/resilient\_dynamic\_estimation}.}. The physical parameters are illustrated in Fig. \ref{fig:invpen}. 
%the The mass of the cart and the mass of the pendulum are both $1$ kilogram. The length of pendulum is $1$ meter and the moment of inertia of the pendulum is $1/3$ $kg\cdot m^2$.
The control input $u(k)$ is the force applied on the cart, and $b_\theta,b_x$ are the friction coefficients at pendulum joint and cart wheels.
The state $x_1,x_2,x_3,x_4$ represent cart position coordinate, cart velocity, pendulum angle from vertical and pendulum angle velocity respectively. 

Consider the system linearized at $x_3=x_4=0$, and we sample the continuous-time linear system periodically with sampling interval $T_s=0.02$ seconds. 
The system equation is:
\begin{align*}
x(k+1)=\begin{bmatrix}
1  &   2.0\cdot 10^{-2}  &  -2.0\cdot 10^{-4}  &  1.9\cdot 10^{-5}\\
0  &   1.0\cdot 10^{0}   & -2.0\cdot 10^{-2}  &   1.8\cdot 10^{-3}\\
0  &   1.0\cdot 10^{-5} &   1.0\cdot 10^{0} &   2.0\cdot 10^{-2}\\
0  &   1.0\cdot 10^{-3}&  2.1\cdot 10^{-1}  &  9.8\cdot 10^{-1}
\end{bmatrix}x(k)+\begin{bmatrix}
2.0\cdot 10^{-4}\\
2.0\cdot 10^{-2}\\
-2.0\cdot 10^{-4}\\
-2.0\cdot 10^{-2}
\end{bmatrix}
u(k)+w(k).
\end{align*}
\begin{align*}
y(k)=
\begin{bmatrix}
1& 0 &0 &0\\
1& 0 &0 &0\\
1& 0 &0 &0\\
0& 0 &1 &0
\end{bmatrix}x(k)+v(k)+a(k).
\end{align*}
The system dynamic matrix can be written as the following Jordan canonical form by an invertible linear transformation:
$$
\begin{bmatrix}
1.057 & 0 &  0 & 0\\
0  & 1 & 0 &  0\\
0  & 0 & 0.999 & 0  \\
0  & 0 & 0 &  0.925 
\end{bmatrix},
$$
and we consider the system after transformation. System matrix $A$ have four Jordan blocks with size $1\times1$ and the upper left two blocks have unstable eigenvalues. Therefore, the set of unstable states and stable states are $\Uc=\{1,2\},\Sc=\{3,4\}.$
The canonical form of $G_i^\Uc$ are 
\begin{align}
H_1^\Uc=H_2^\Uc=H_3^\Uc=\begin{bmatrix}
1 & 0 \\
0  & 1 \\
0  & 0   \\
0  & 0 
\end{bmatrix},\ H_4^\Uc=\begin{bmatrix}
0 & 0 \\
0  & 1 \\
0  & 0   \\
0  & 0 
\end{bmatrix}.
\end{align}
Only the first 3 sensors can observe unstable state 1, i.e., $\Ec_1=\{1,2,3\}$. All the four sensors can observe unstable state 2, i.e., $\Ec_2=\{1,2,3,4\}$.
Therefore, the system is 2-sparse detectable and our proposed estimator is secure in the presence of 1 corrupted sensor. 
In the simulation, the noise covariances of the system are $Q=R=T_s^2\times\diag(0.1,0.1,0.01,0.01)$.
The initial state is assumed to be known by the estimator.
The controller of the system is designed as a Linear-Quadratic Regulator (LQR), and the feedback matrix is chosen as $K_{\rm lqr}=\begin{bmatrix}
-8& -15& -115 &-32
\end{bmatrix}.
$
%
%
%In the following we perform state transformation on the origin system to better analysis the observability structure.
%There exists a invertible matrix $U$ such that 
%
%In the following, we study the following system where $\bar{C}\triangleq CU$:
%	\begin{align*}
%		\bar{x}(k+1)&=\bar{A} \bar{x}(k)+U^{-1}w(k)+U^{-1}Bu(k), \\
%		y(k)&=\bar{C} \bar{x}(k)+v(k)+a(k).
%	\end{align*}

\begin{figure}[htpb!]
	\centering
	\input{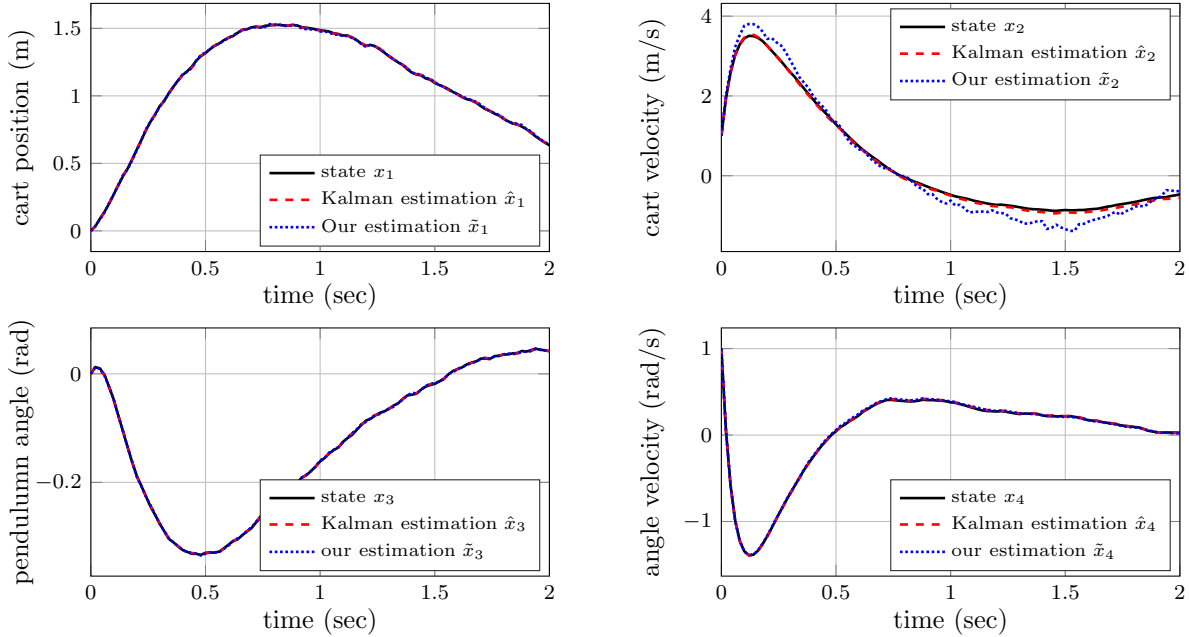}
	\caption{Estimation of states in the absence of attack. The initial state is $x(0)=[0,\ 1,\ 0,\ 1]{'}$. } \label{fig:close_loop}
\end{figure}

We first illustrate the performance of estimation on close-loop system where $u(k)=-K_{\rm lqr} x(k)$.
Fig. \ref{fig:close_loop} presents the performance of the estimation of system states in the absence of attack. Our proposed estimation substantially coincides with the Kalman estimation. The numerical difference attributes to large Gaussian noise that occurs occasionally which violates inequality \eqref{eq:kalman_cond} and error in numerical calculation.

\begin{figure}[ht]
	\centering
	\input{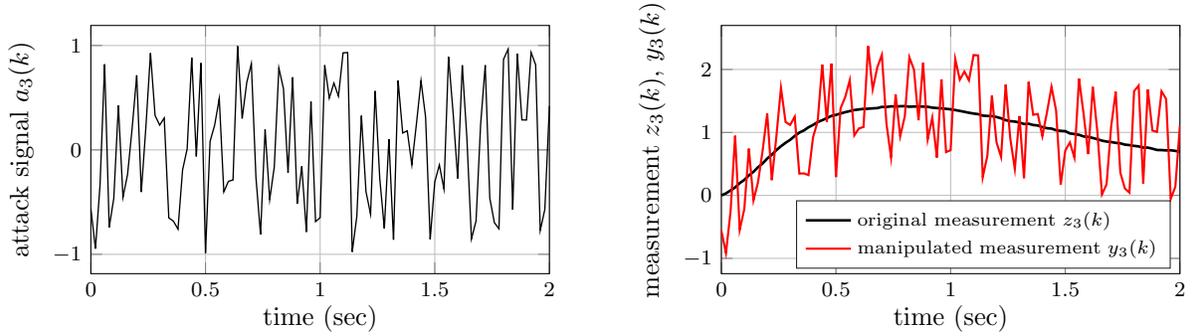}
	\caption{Attack signal and measurements of sensor 3. The attack signal is uniformly distributed in interval $(-1,1)$.  } \label{fig:attack_signal}
\end{figure}

\begin{figure}[ht]
	\centering
	\input{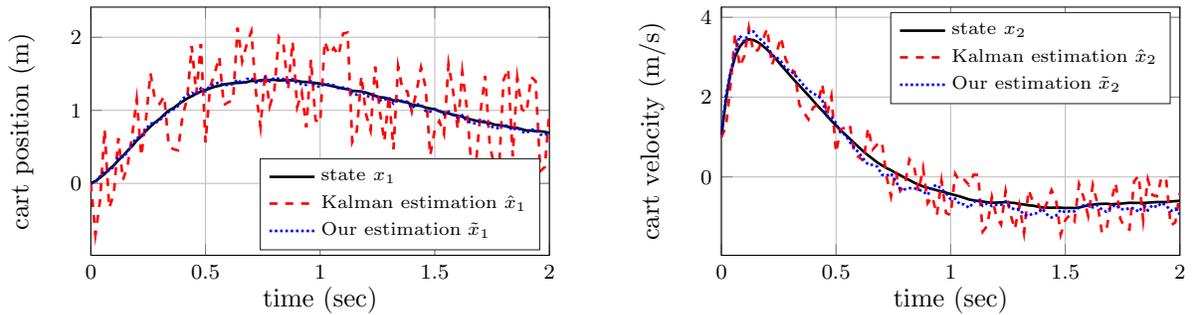}
	\caption{Estimation of states under attack on sensor 3. The attack signal is uniformly distributed in interval $(-1,1)$.  } \label{fig:close_loop_attack}
\end{figure}

Fig. \ref{fig:attack_signal} shows the injected attack signal on sensor 3 and the corresponding observations from sensor 3.
The attack $a_3(k)$ is a time-independent random value uniformly distributed on interval $(-1,1)$. Fig. \ref{fig:close_loop_attack} demonstrates the estimation with the attack shown in Fig. \ref{fig:attack_signal}.
As shown in the Fig. \ref{fig:close_loop_attack}, Kalman estimation (denoted as red dashed line) has larger estimation error than our proposed estimation under the attack.

Fig. \ref{fig:MSE} illustrates the estimation mean square error (MSE$=1/N\sum_{k=1}^{N}\left\|\hat{x}(k)-x(k)\right\|_2^2$) of our proposed estimator with varying tuning parameter $\gamma$ and varying attack magnitude with sensor 3 corrupted. 
The number of time steps is set as $N=200$.
The attack signal $a_3(k)$ is uniformly distributed in interval $(-\|a\|_\ift,\|a\|_\ift)$.
In Fig. \ref{fig:MSE_gamma}, the MSE of the oracle Kalman estimation is illustrated by the red dashed line and that of the Kalman estiamtion under attack is illustrated by the red solid line.
As shown in Fig. \ref{fig:MSE_gamma}, by properly choosing $\gamma$, the MSE of our proposed estimator is smaller than that of Kalman estimation (blue line is below the red horizontal line), with the cost that MSE without attack being slightly larger. 
In Fig. \ref{fig:MSE_mag}, as the magnitude of attack signal increases, MSE of Kalman estimator increases significantly while our proposed secure estimator holds low MSE despite increasing magnitude of injected attack signal.

\begin{figure}[ht]
	\centering
	\begin{subfigure}{.48\textwidth}
		\centering
		\begin{tikzpicture}
\begin{axis}
[	width=2.5in,
	height=1.5in,
	at={(0in,0in)},
	scale only axis,
	xmin=-0.01,
	xmax=0.2,
	ymin=-0.02,
	ymax=0.5,
	title={MSE with varying $\gamma$},
	xlabel={MSE without attack},
	ylabel={MSE under attack},
	x label style={at={(axis description cs:0.5,0.0)},anchor=north},
	y label style={at={(axis description cs:0.1,.5)},anchor=south},
	xticklabel style = {font=\footnotesize},
	yticklabel style = {font=\footnotesize},
	axis background/.style={fill=white},
	xmajorgrids,
	ymajorgrids,
	legend style={at={(0.98,0.98)}, anchor=north east, legend cell align=left, align=left, draw=white!15!black, font=\scriptsize}]
	
	 \addplot [color={blue}, line width=1pt, mark=*, mark size=1pt]
	table[row sep={\\}]
	{
%		2.73294  2.18893 \\ % gamma=0.1
%		2.7611  2.2588  \\  % gamma=0.2
%		2.73456  2.14272 \\ % gamma=0.5
%		2.70273  2.23093 \\ % gamma=1
%		2.5806  2.13751 \\  % gamma=2
%		 1.3291898141799852 \\ % gamma=4
		0.49171288806410334  0.23058873634922077 \\ % gamma=5
		0.15031158033658748 0.023028041755303673 \\% gamma=6
		0.022956004079827286  0.023069623748774245 \\ % gamma=7
		0.022892999120518853  0.02302036422425081 \\ % gamma=8
		0.022738740120155414  0.023028041755303673 \\ % gamma=10
		0.02031379510833805 0.023522751164769238 \\ % gamma=20
		0.018767389239823917 0.025297232039458894   \\ % gamma=30
		0.018239965010654734 0.03540481351322025 \\ % gamma=40
		0.017382121611668622 0.053386825865487555\\ % gamma=50
		0.015534591302800367 0.06477553708429706\\ % gamma=100
		0.015363288434118163 0.16504007534811177\\ % gamma=200
		0.015363288434118163 0.27504007534811177\\ % gamma=500
		0.012899244187193867 0.382451394009979\\ % gamma=1000
	} 
	;\addlegendentry{Our proposed estimator}

	 \addplot [dashed, color={red}, line width=1pt]
	table[row sep={\\}]
	{
	 3  0.0016205180878490848  \\
	 0.0016205180878490848  0.0016205180878490848 \\
	 0.0016205180878490848  30\\
	}
	;\addlegendentry{Oracle Kalman estimator}
	
	 \addplot [color={red}, line width=1pt]
	table[row sep={\\}]
	{
		0.002  0.3483651642262539\\
		3  0.3483651642262539\\
	}
	;%\addlegendentry{Kalman estimator}
	
	\draw[latex-] (2.7, 1)--(4,2) node[right]{  $\gamma=100$};
	\draw[latex-] (3.3, 0.6)--(4,1.3) node[right]{ $\gamma=7$};
	\draw[-] (6, 3.65)--(12,3) node[below=-3pt]{\small MSE of Kalman under attack};

\end{axis}

\end{tikzpicture}
		\caption{\normalsize  Estimation mean square error (MSE) with varying tuning parameter $\gamma$. The attack magnitude is $\|a\|_\ift=1$.  } \label{fig:MSE_gamma}
	\end{subfigure}
	\hspace{5pt}
	\begin{subfigure}{.48\textwidth}
		\centering
		\begin{tikzpicture}

\begin{axis}
	[width=2.5in,
	height=1.5in,
	at={(3.5in,0in)},
	scale only axis,
	xmin=0,
	xmax=2.0,
	ymin=-0.1,
	ymax=1.5,
	title={MSE with varying attack magnitude},
	xlabel={attack magnitude $||a||_\ift$},
	ylabel={MSE under attack},
	x label style={at={(axis description cs:0.5,0)},anchor=north},
	y label style={at={(axis description cs:0.1,.5)},anchor=south},
	xticklabel style = {font=\footnotesize},
	yticklabel style = {font=\footnotesize},
	axis background/.style={fill=white},
	xmajorgrids,
	ymajorgrids,
	legend style={at={(0.02,0.98)}, anchor=north west, legend cell align=left, align=left, draw=white!15!black, font=\scriptsize}]

		\addplot [color={blue}, line width=1pt, mark=*, mark size=1pt]
		table[row sep={\\}]
		{   0 0.022738740120155414\\
			0.1 0.02273998743284275\\
			0.2 0.02309437925527503\\
			0.5 0.022727416340979578\\
			1 0.023028041755303673\\
			1.5 0.04150262265679778\\
			2  0.06361155341192473\\
		}
		;\addlegendentry{Our proposed estimator}
		
		\addplot [color={red}, line width=1pt, mark=*, mark size=1pt]
		table[row sep={\\}]
		{   0 0.0016205180878490848\\
			0.1 0.005836883178707939\\
			0.2 0.016944741009078466 \\
			0.5 0.09868982450677061 \\
			1   0.3483651642262539\\
			1.5 0.7777564157424298\\
			2 1.394430537838036 \\
		}
		;\addlegendentry{Kalman estimator}

%	\draw[latex-] (0.2,24)--(0.4,31) node[right]{\footnotesize $\gamma=100$};
%	\draw[latex-] (0.42,13.4)--(1,13) node[right]{\footnotesize $\gamma=30$};
%	\draw[latex-] (2,8)--(2.4,10) node[right]{\footnotesize $\gamma=5$};
%	\draw[-] (1.4, 17.6)--(1.6,21) node[above]{\footnotesize MSE of Kalman under attack};
	
\end{axis}

\end{tikzpicture}
		\caption{\normalsize Estimation mean square error (MSE) with varying attack magnitude. The parameter $\gamma$ is set as $\gamma=10$. } \label{fig:MSE_mag}
	\end{subfigure}
	\caption{}
	\label{fig:MSE}
\end{figure}

\section{Conclusion}\label{sec:conclusion}
This paper considers LTI system with Gaussian noise against sparse integrity attack on a subset of sensors. 
Under the geometric multiplicity assumption on stable eigenvalues of $A$, we propose an estimation scheme that is secure to $(p,m)$-sparse attack as long as the system is $2p$-sparse detectable. 
To achieve this, we first prove that the span of the rows of $G^\Uc_i$ is equivalent to the observable space corresponding to unstable states, based on which the canonical form $H_i$ is designed. The proposed estimator is formulated as a convex optimization problem based on $H_i$, where by careful design, the estimation of stable states is always secured. 
Moreover, in the absence of attack, the proposed estimation coincides with Kalman estimation for certain probability, which can be adjusted by tuning parameter $\gamma$ to balance between the performance with and without attack.
We further prove that the $2p$-sparse detectability is the fundamental limit of secure dynamic estimation.
Our proposed estimator achieves this fundamental limit with good performance in the absence of attack and low computation complexity.

%%%%%%%%%%%%%%%%%%%%%%%%%%%%%%%%%%%%%%%%%%%%%%%%%%%%%%%%%%%%%%%%%%%%%%%%%%%%%%%%

%\backmatter

%\section*{Acknowledgments}
%This is acknowledgment text.\cite{Kenamond2013} Provide text here. This is acknowledgment text. Provide text here. This is acknowledgment text. Provide text here. This is acknowledgment text. Provide text here. This is acknowledgment text. Provide text here. This is acknowledgment text. Provide text here. This is acknowledgment text. Provide text here. This is acknowledgment text. Provide text here. This is acknowledgment text. Provide text here. 

%\subsection*{Author contributions}
%
%This is an author contribution text. This is an author contribution text. This is an author contribution text. This is an author contribution text. This is an author contribution text. 
%
%\subsection*{Financial disclosure}
%
%None reported.
%
%\subsection*{Conflict of interest}
%
%The authors declare no potential conflict of interests.

%\section*{Supporting information}
%
%The following supporting information is available as part of the online article:
%
%\noindent
%\textbf{Figure S1.}
%{500{\uns}hPa geopotential anomalies for GC2C calculated against the ERA Interim reanalysis. The period is 1989--2008.}
%
%\noindent
%\textbf{Figure S2.}
%{The SST anomalies for GC2C calculated against the observations (OIsst).}

\appendix
\section*{Appendix}
%\textbf{\Large Appendix}
\section{Proof of Theorem 2}\label{ap:span}
\begin{proof}
	Define the characteristic polynomial of $A$ as $p(x)=a_n x^n +\cdots+a_1 x +a_0$.
	Define polynomial fraction  $q_\pi(x)$ with respect to constant $\pi$ as
	$q_\pi(x)=\frac{p(x)-p(\pi)}{x-\pi}$ where $x\neq \pi$.
	Therefore,
	$$q_\pi(A)(A-\pi I) = p(A)-p(\pi)I=-p(\pi)I ,$$
	where the last equality comes from Cayley-Hamilton Theorem.
	As a result, when $\pi$ is not the eigenvalue of $A$, we have
	\begin{align}\label{A-lambdaI}
	(A-\pi I)^{-1}=-\frac{1}{p(\pi)} q_\pi(A).
	\end{align}
	In order to simplify notations, we define 
	\begin{equation}\label{eq:bjk}
	b_{j,k}\triangleq-\frac{1}{p(\pi_j)}\sum_{i=0}^{n-k-1} a_{i+k+1} \pi_j^i,
	\end{equation}
	where $\pi_j$ is the $j$-th diagonal element of $\Pi$, i.e., $j$-th eigenvalue of $A-KCA$ as defined in \eqref{eq:VLambda}.
	According to (\ref{A-lambdaI}), the $j$-th row of matrix $G_i$ can be reformulated as
	$$C_{i} A\left(A-\pi_{j} I\right)^{-1}=
	\begin{bmatrix}
	b_{j,0} & b_{j,1} & \cdots  & b_{j,n-1} 
	\end{bmatrix} O_i A.$$
	Therefore, $G_i$ can be interpreted as follows
	\begin{align*}
	G_i = \begin{bmatrix}
	b_{1,0} & b_{1,1} & \cdots  & b_{1,n-1} \\
	b_{2,0} & b_{2,1} & \cdots  & b_{2,n-1} \\
	\vdots & \vdots & \ddots  & \vdots \\
	b_{n,0} & b_{n,1} & \cdots  & b_{n,n-1} 
	\end{bmatrix}
	O_i A 
	= & \mathcal{D}_1\mathcal{D}_2\mathcal{D}_3 O_i A , %\label{eq:GandOA}		
	\end{align*}
	where $\mathcal{D}_1\triangleq\text{diag}\left(-\frac{1}{p(\pi_1)},-\frac{1}{p(\pi_2)},\cdots,-\frac{1}{p(\pi_n)}\right)$,
	\begin{align*}
	\mathcal{D}_2\triangleq
	\begin{bmatrix}
	\pi_1^{n-1} & \pi_1^{n-2} & \cdots  & 1 \\
	\pi_2^{n-1} & \pi_2^{n-2} & \cdots  & 1 \\
	\vdots & \vdots & \cdots  & \vdots \\
	\pi_n^{n-1} & \pi_n^{n-2} & \cdots  & 1
	\end{bmatrix}, 
	\mathcal{D}_3\triangleq
	\begin{bmatrix}
	a_n & 0 & \cdots &   0 \\
	a_{n-1} & a_n & \cdots &   0 \\
	\vdots & \vdots & \ddots  & \vdots \\
	a_1 & a_2 & \cdots  & a_n 
	\end{bmatrix}.
	\end{align*}
	According to Assumption \ref{as:distinct_eigvalue}, all $\pi_j$ are distinct eigenvalues and they are not the eigenvalues of $A$, i.e. the diagonal matrix $\mathcal{D}_1$ and the Vandermonde matrix $\mathcal{D}_2$ are invertible. Moreover, $a_n=1$. Therefore, the lower triangular Toeplitz matrix $\mathcal{D}_3$ is invertible and thus $\rs(G_i)=\rs(O_i A)$. 		
	We continue to prove $\rs(O_i)=\rs(O_i A)$. Considering that $A^n=-a_{n-1}A^{n-1}-\cdots-a_0 I$, one obtains the following equation \eqref{eq:O_OA}.
	\begin{equation}
	\label{eq:O_OA}
	O_i A=
	\begin{bmatrix}
	0 & 1 & 0 &  \cdots & 0 \\
	0 & 0 & 1 &  \cdots & 0 \\
	\vdots & \vdots & \vdots & \ddots & \vdots \\
	0 & 0 & 0 &  \cdots & 1 \\
	-a_0 & -a_1 & -a_2 & \cdots &  -a_{n-1}
	\end{bmatrix}
	O_i .
	\end{equation}	%\vspace{-20pt}
	According to Assumption \ref{as:distinct_eigvalue}, $A$ is invertible and $a_0=(-1)^n\det(A)\neq 0$, which leads to the equation that $\rs(O_i)=\rs(O_i A)$.
	As a result, $\rs(G_i^\Uc)$=$\rs(O_i^\Uc)$. We continue to prove that $\rs(O_i^\Uc)=\rs(H_i^\Uc).$
	Since $A_1$ is assumed to be in the Jordan canonical form and all eigenvalues have geometric multiplicity 1, one can verify that nonzero columns of $O^\Uc_i$ are linear independent. Therefore, $i\in\Ec_j$ is equivalent to that $j$-th column of $O_i$ is non-zero, i.e., $O^\Uc_i$ has the same row-span with the canonical form $H^\Uc_i$.
	
\end{proof}

%\section{Proof of Theorem \ref{th:att_obs}}\label{ap:att_obs}
%
%\begin{proof}[Proof of Theorem \ref{th:att_obs}]
%	In view of Theorem \ref{th:TAC} and Lemma \ref{lm:least_square}, it suffices to prove that the following two propositions are equivalent:
%	\begin{enumerate}
%		\item The system is $2p$-sparse observable.
%		\item The following inequality holds for all $x \neq \mathbf{0}$, $x\in\Rb^n$:
%		\begin{equation}\label{eq:proof_cond}
%		\sum_{i \in \mathcal{I}}\left\|H_{i} x\right\|_{1}<\sum_{i \in \mathcal{I}^{c}}\left\|H_{i} x\right\|_{1}, \quad \forall\ \Ic\subset \Jc, |\mathcal{I}|\leq p .
%		\end{equation}
%	\end{enumerate}
%	Based on the form of $H_i$ in Theorem \ref{th:span}, inequality \eqref{eq:proof_cond} can be written as 
%	\begin{equation*}
%	\sum_{j=1}^{n} \sum_{i \in \mathcal{I}\cap\Ec_j}|x_j|<\sum_{j=1}^{n} \sum_{i \in \mathcal{I}^c\cap\Ec_j}|x_j| 
%	\end{equation*}
%	or equivalently 
%	$$
%	\sum_{j=1}^{n} c_j(\Ic)\cdot|x_j|<\sum_{j=1}^{n} h_j(\Ic)\cdot|x_j| .
%	$$
%	If the system is $2p$-sparse observable, we have $c_j(\Ic)<h_j(\Ic)$ for all $j\in\{1,2,\cdots,n\}$. Thus, $\sum_{j=1}^{n} \left(h_j(\Ic)-c_j(\Ic)\right)\cdot|x_j| >0$ holds for all $x\neq \mathbf{0}$.
%	If the system is not $2p$-sparse observable, there exists $\Ic^*$ with $|\Ic^*|=p$ and $j^*$ such that $c_{j^*}(\Ic^*)>h_{j^*}(\Ic^*)$. One can design $x\neq \mathbf{0}$ such that 
%	$$\left(c_{j^*}(\Ic^*)-h_{j^*}(\Ic^*)\right)\cdot|x_{j^*}|> \sum_{j\neq j^*}\left(h_{j}(\Ic^*)-c_{j}(\Ic^*)\right)\cdot|x_j| .$$
%	Therefore, condition \eqref{eq:proof_cond} is violated and the proof is completed. 
%\end{proof}

\section{Proof of Theorem 3}\label{ap:ap_no_attack}
\begin{proof}[Proof of Theorem \ref{th:no_attack}]
	
	Considering the KKT condition of problem \eqref{pb:resilient_LASSO}, one obtains that if 
	$$\left\|\Wc \begin{bmatrix}
	\mu(k) \\
	N H \tilde{x}(k)
	\end{bmatrix}\right\|_\ift\leq\gamma,$$ then the solution $\nu(k)$ satisfy that $\nu(k)=\mathbf{0}$. In this scenario, solutions to problem \eqref{pb:resilient_LASSO} and problem \eqref{pb:least_square} are equivalent and the solution $\tilde{x}(k),\mu(k),\nu(k)$ satisfy
	\begin{equation}\label{eq:lasso_recover}
	\tilde{x}(k)=\tilde{x}_\ls(k)=\hat{x}(k),\ \mu(k)=\varphi(k),\ \nu(k)=\mathbf{0}.
	\end{equation}
	According to Lemma \ref{lm:least_square}, the solution $\varphi(k)$ of problem \eqref{pb:least_square} satisfy the following equation:
	\begin{equation}\label{eq:ls_recover}
	\tilde{x}_\ls(k)=\hat{x}(k),\ \varphi(k)=(I-GF)\epsilon(k),
	\end{equation}
	where $\hat{x}(k)$ is the fixed gain Kalman estimation defined in \eqref{eq:fix_gain_kalman}.
	Combining \eqref{eq:ls_recover} and \eqref{eq:lasso_recover}, result in Theorem \ref{th:no_attack} is obtained. 
\end{proof}

\section{Proof of Theorem 4}\label{ap:main}
Before proving Theorem \ref{th:main}, we need the following Lemma.
Define the number of honest sensors and compromised sensors (w.r.t. compromised set $\Ic$) that can observe state $j$ as:
\begin{align*}
h_j(\Ic)\triangleq |\Ec_j\cap \Ic^c|, \
c_j(\Ic)\triangleq |\Ec_j\cap \Ic|.
\end{align*}
We have the following lemma quantifying the property of $h_j(\Ic)$ and $c_j(\Ic)$.  
\begin{lemma}\label{lm:cj<hj}
	The following two propositions are equivalent.
	\begin{enumerate}
		\item The system is $2p$-sparse observable.
		\item For any $\Ic$ with $|\Ic|=p$, the inequality $c_j(\Ic)<h_j(\Ic)$ holds for all $j\in\{1,2,\cdots,n\}$.
	\end{enumerate}
\end{lemma}
\begin{proof}[Proof of Lemma \ref{lm:cj<hj}]
	We prove the contrapositive of (1)$\Rightarrow$(2). Supposing that there exists $j^*$ and $\Ic^*$ with $|\Ic^*|=p$ such that $ c_{j^*}(\Ic^*)\geq h_{j^*}(\Ic^*)$, then
	$h_{j^*}(\Ic^*)\leq c_{j^*}(\Ic^*)\leq |\Ic^*|=p$. Noticing that $c_j(\Ic)+h_j(\Ic)=|\Ec_j|$ holds for all $\Ic$, we have $|\Ec_{j^*}|\leq 2p$. 
	There exists set $\Ac$ that satisfy $\Ac\supseteq\Ec_{j^*}$ and $|\Ac|=2p$.
	According to the definition of $\Ec_{j^*}$, there exists no sensor in set $\Jc\setminus \Ac$ who can observe state $j^*$, i.e.,
	\begin{equation*}
	e_{j^*}\notin \rs(O_i),\ \forall i\in\Jc\setminus \Ac.
	\end{equation*}
	As a result, system $(A,C_{\Jc\setminus\Ac})$ is not $2p$-sparse observable according to Definition \ref{df:sparse_obs}. 
	
	We proceed to prove (2)$\Rightarrow$(1). 
	Since for any $\Ic$ with $|\Ic|=p$, $h_j(\Ic)>c_j(\Ic)\geq 0$, the system sparse observability index is at least $p$. Therefore, for each $j\in\{1,2,\cdots,n\}$, there exists an $\Ic^*$ such that $c_j(\Ic^*)=p$, and thus $|\Ec_j|=h_j(\Ic^*)+c_j(\Ic^*)\geq 2p+1$. According to the definition $\Ec_j$, there are at least $2p+1$ sensors that can observe sensor $j$, and the system is $2p$-sparse observable.	
	
\end{proof}

We need the following notations for the proof.
Define the unstable part and stable parts of $x$ as the following where $x_u\in\Cb^{n_u\times 1}$ and $x_s\in\Cb^{n_s\times 1}$. Similarly, divide matrix $H_i$ into four parts based on Theorem \ref{th:span} where $H_{uu,i}\in\Cb^{n_u\times n_u}$ and $H_{ss,i}\in\Cb^{n_s\times n_s}$.
\begin{align*}
x = 
\left[
\begin{array}{c}
x_u\\x_s
\end{array}
\right],\
%\begin{array}{l}
%\hspace{-6pt}\left. \right\} \text{first $n_u$ entries}\\
%\hspace{-6pt}\left. \right\} \text{last $n_s$ entries}
%\end{array} .\\
H_i=\begin{bmatrix}
H_{uu,i}	& 	H_{us,i} \\
\mathbf{0}_{n_s\times n_u} & H_{ss,i}
\end{bmatrix} .
\end{align*}
Define $\eta_{i}\triangleq P_i\zeta_{i}$.
Similar to $x$, $\tilde{x}$ and $\eta_i$ are also divided to $\tilde{x}_u,\tilde{x}_s,\eta_{i,u},\eta_{i,s}$ in the same way.

\begin{proof}[Proof of Theorem \ref{th:main}]
	Consider the KKT condition of problem (\ref{pb:resilient_LASSO}) and denote the dual variables for equation constraints as $\lambda=[\lambda_1{'},\cdots,\lambda_m{'}]{'}\in \Cb^{mn\times 1}$:	
	\begin{align}
	(\tilde{M}^{-1}+N{'}N)\mu+N{'} N H\tilde{x} - \lambda &= \mathbf{0}, \label{eq:KKT1}\\
	H{'} N{'}N\mu+H{'} N{'} N H\tilde{x} -  H{'}\lambda &= \mathbf{0}, \label{eq:KKT2} \\
	\gamma \cdot  \tilde{\nu} - \lambda &= \mathbf{0}, \label{eq:KKT3} \\
	{Y} - H \tilde{x} - \mu - \nu &= \mathbf{0},  \label{eq:KKT4}
	\end{align}
	where $\tilde{\nu}$ belongs to the subgradient of $\|\nu\|_1$, i.e., for the $i$-th entry:
	\begin{align*}
	\begin{cases}
	[\tilde{\nu}]_i=-1 ,& \ [\nu]_i<0 \\
	[\tilde{\nu}]_i=1 , &\ [\nu]_i>0  \\
	[\tilde{\nu}]_i\in\left[-1,1\right],& \ [\nu]_i=0 
	\end{cases}.
	\end{align*}
	
	Combining (\ref{eq:KKT1}) and (\ref{eq:KKT2}) leads to:
	\begin{equation}\label{eq:KKT12}
	\begin{bmatrix}
	\tilde{M}^{-1}+N{'}N & N{'} N H\\
	H{'} N{'} N  & H{'} N{'} N H
	\end{bmatrix}
	\begin{bmatrix}
	\mu \\ \tilde{x}
	\end{bmatrix}=
	\begin{bmatrix}
	\lambda \\ H{'} \lambda 
	\end{bmatrix}.
	\end{equation}
	According to the definition of $N$, the first $n_u$ rows of $H{'} N{'} N$ are zeros. Therefore, we extract the non-zeros part of equation (\ref{eq:KKT12}) in the following:
	\begin{equation}\label{eq:KKT12_nonzero}
	\begin{bmatrix}
	\tilde{M}^{-1}+N{'}N & N{'} N H \Lc{'}\\
	\Lc H{'} N{'} N  &  \Lc H{'} N{'} N H \Lc{'}
	\end{bmatrix}
	\begin{bmatrix}
	\mu \\ \tilde{x}_s
	\end{bmatrix}=
	\begin{bmatrix}
	\lambda \\ \Lc H{'} \lambda 
	\end{bmatrix}.
	\end{equation}
	where $\Lc\triangleq 
	\begin{bmatrix}
	\mathbf{0}_{n_s\times n_u} & I_{n_s}
	\end{bmatrix}.
	$
	%Notice that
	%\begin{align}
	%&\Lc H{'} N{'} N H \Lc{'} 
	%- \Lc H{'} N{'} N(\tilde{M}^{-1}+N{'}N)^{-1} N{'} N H \Lc{'} \notag \\
	%= &\Lc H{'} N{'}  \left(I -N(\tilde{M}^{-1}+N{'}N)^{-1}N{'} \right)  N H \Lc{'} . \label{eq:schur_comp}
	%\end{align}
	%Since I−N(M~−1+N⊤N)−1N⊤ is invertible and \LcH⊤N⊤ is row full-rank (???) is also invertible due to the Frobenius rank inequality.
	%Therefore, according to the result of Schur complement, matrix on the left of (???) is invertible.
	Rewrite (\ref{eq:KKT12_nonzero}) as:
	\begin{align}\label{eq:KKT12_matrix}
	\left(
	\begin{bmatrix}
	I_{mn} & \mathbf{0} \\
	\mathbf{0}  &  \Lc H{'} N{'}
	\end{bmatrix}
	\Wc
	\begin{bmatrix}
	I_{mn} & \mathbf{0} \\
	\mathbf{0}  &  N H \Lc{'}
	\end{bmatrix}
	\right)
	\begin{bmatrix}
	\mu \\ \tilde{x}_s
	\end{bmatrix}=
	\begin{bmatrix}
	I_{mn} & \mathbf{0} \\
	\mathbf{0}  &  \Lc H{'}
	\end{bmatrix}
	\lambda .
	\end{align}
	Notice that $\Wc$ is positive definite and $\begin{bmatrix}
	I_{mn} & \mathbf{0} \\
	\mathbf{0}  &  \Lc H{'} N{'}
	\end{bmatrix}$
	is full row-rank,
	%	\footnote{
	%		$\Lc H{'} N{'}=
	%		\begin{bmatrix}
	%		H{'}_{ss,1} & H{'}_{ss,2} & \cdots & H{'}_{ss,m} % !!!NOtice that H_{ss,i}=G_{ss,i}, if this matrix not full row fank, then $G{'}_{ss,1} & G{'}_{ss,2}$ is row rank defected and there are unobservable stable states.
	%		\end{bmatrix}
	%		$ is row full-rank due to the fact that for every stable state, there is at least one sensor who is able to observe the state.
	%	}, 
	due to the Frobenius rank inequality, the matrix on the left of (\ref{eq:KKT12_matrix}) is also invertible, and thus the following matrix is well-defined:
	\begin{align}\label{eq:def_H}
	\Fc	\triangleq
	\left(
	\begin{bmatrix}
	I_{mn} & \mathbf{0} \\
	\mathbf{0}  &  \Lc H{'} N{'}
	\end{bmatrix}
	\Wc
	\begin{bmatrix}
	I_{mn} & \mathbf{0} \\
	\mathbf{0}  &  N H \Lc{'}
	\end{bmatrix}
	\right)^{-1}
	\begin{bmatrix}
	I_{mn} & \mathbf{0} \\ \mathbf{0} &\Lc H{'}
	\end{bmatrix}.
	\end{align}
	
	%Sufficient condition of (???) is:
	%\begin{align}\label{eq:KKT2_2}
	%	N \mu + N H \tilde{x} - \frac{1}{2}(N^\dag){'} \lambda = \mathbf{0} 
	%\end{align}
	%where N\dag is the pseudo-inverse of N.
	%Combining (???) and (???) leads to:
	%\begin{equation}\label{eq:KKT12}
	%	\begin{bmatrix}
	%		\tilde{M}^{-1}+N{'}N & N{'} \\
	%		N & I
	%	\end{bmatrix}
	%	\begin{bmatrix}
	%		\mu \\ N H \tilde{x}
	%	\end{bmatrix}=\frac{1}{2}
	%	\begin{bmatrix}
	%		\lambda \\ (N^\dag){'} \lambda 
	%	\end{bmatrix}.
	%\end{equation}
	
	According to (\ref{eq:KKT3}), $\|\lambda\|_\infty \leq \gamma$. 
	Therefore we have the following from (\ref{eq:KKT12_matrix})
	\begin{equation}\label{eq:mu_xs_bounded}
	\left\|\begin{bmatrix}
	\mu \\ \tilde{x}_s 
	\end{bmatrix}\right\|_\infty
	\leq \gamma \cdot
	\left\|\Fc\right\|_\infty.
	\end{equation}
	
	Now we continue to prove that the estimation of unstable states $\tilde{x}_u$ are resilient.
	Rewrite the optimization problem \eqref{pb:resilient_LASSO} as 
	\begin{align*}
	\underset{{\tilde{x}},\ \mu}{\text{minimize}}&\quad \frac{1}{2} 
	\begin{bmatrix}
	\mu \\
	N H \tilde{x}
	\end{bmatrix}^{'} \Wc
	\begin{bmatrix}
	\mu \\
	N H \tilde{x}
	\end{bmatrix} + \gamma\left\|{Y}-\mu-H \tilde{x}\right\|_1 
	\end{align*}
	where the time index is omitted for notation simplicity.
	Consider the 1-norm term in the objective function:
	\begin{align*}
	&\left\| {Y}-\mu-H \tilde{x} \right\|_1
	=\sum_{i=1}^{m} 
	\left\|\eta_{i,u}-\mu_{i,u}-(H_{uu,i} \tilde{x}_u + H_{us,i} \tilde{x}_s) \right\|_1 
	+ \sum_{i=1}^{m} \left\|\eta_{i,s}-\mu_{i,s}- H_{ss,i} \tilde{x}_s \right\|_1 
	\end{align*}
	where $\eta_{i,u}, \mu_{i,u}$ is the vector composed of first $n_u$ element of $\eta_{i}, \mu_{i}$ and 
	$\eta_{i,s}, \mu_{i,s}$ is the vector composed of last $n_s$ element of $\eta_{i}, \mu_{i}$.
	Suppose that $\mu$ and $\tilde{x}_s$ have taken the value of optimal solution $\mu^*, \tilde{x}_s^*$, it is sufficient to minimize the following :
	\begin{align}\label{pb:1-norm_min}
	\min_{\tilde{x}_u} \sum_{i=1}^{m} \left\|\eta_{i,u}-\mu^*_{i,u}- H_{us,i} \tilde{x}^*_s - H_{uu,i} \tilde{x}_u \right\|_1  .
	\end{align}
	Define $\xi_i\triangleq \eta_{i,u}-\mu^*_{i,u}- H_{us,i} \tilde{x}^*_s $ and recall $[\cdot]_j$ is the $j$-th entry of a vector. The objective function in (\ref{pb:1-norm_min}) can be written as 
	\begin{align}
	&\sum_{i=1}^{m}\sum_{j=1}^{n_u} \left| [\xi_i]_j -[H_{uu,i} \tilde{x}_u]_j \right| 
	=\sum_{j=1}^{n_u}\sum_{i\in\Ec_j} \left| [\xi_i]_j -\tilde{x}_j \right| . \label{eq:solve_for_median}
	\end{align}
	where $\Ec_j$ is the index set of sensors that can observe state $j$ that is defined in \eqref{eq:def_Ec}.
	For each unstable state $j\in\Uc$, the minimizer $\tilde{x}_j$ of objective (\ref{eq:solve_for_median}) could be explicitly written as the median of all $[\xi_i]_j$ among $i\in\Ec_j$.
	
	Before proving that $\tilde{x}_j$ is bounded, let us define the following operator: $f_{i}: R \times R \times \cdots \times R \rightarrow R,$ such that $f_{i}\left(\alpha_{l}, l\in\{1,\cdots,L\}\right)$ equals to the $i$-th smallest element in the set $\left\{\alpha_{1}, \ldots, \alpha_{L}\right\} .$ For even number $i$, we further define 
	$$f_{\frac{i+1}{2}} = \left(f_{\frac{i}{2}} + f_{\frac{i}{2}+1}\right)/2.$$ 
	Thus, $f_{(L+1)/2}\left(\alpha_{l}, l\in\{1,\cdots,L\}\right)$ is the median number of set $\left\{\alpha_{1}, \ldots, \alpha_{L}\right\}$ and the solution to problem \eqref{pb:1-norm_min} is 
	\begin{align*}
	\tilde{x}_j = f_{(|\Ec_j|+1)/2}\left([\xi_i]_j, i\in\Ec_j\right),j\in\Uc .
	\end{align*}
	Define the uncorrupted data corresponding to sensor $i$ as $\eta^o_{i}=P_i\zeta^o_i$. Define $\xi^o_{i}$ correspondingly as $\xi^o_i\triangleq \eta^o_{i,u}-\mu^*_{i,u}- H_{us,i} x^*_s $.
	Recalling that the number of honest sensors and compromised sensors that can observe unstable state $j\in\Uc$ are $h_j$ and $c_j$, we have
	\begin{align}	
	f_{(h_j-c_j)}\left([\xi^o_i]_j, i\in\Ec_j\right) &\leq
	f_{(m+1)/2}\left([\xi_i]_j, i\in\Ec_j\right), \label{eq:x_u_leftbound}\\
	f_{(m+1)/2}\left([\xi_i]_j, i\in\Ec_j\right)&\leq 
	f_{2c_j}\left([\xi^o_i]_j, i\in\Ec_j\right) .  \label{eq:x_u_rightbound}
	\end{align}\	According to Lemma \ref{lm:cj<hj}, $h_j-c_j>0$ and $2c_j<h_j+c_j=|\Ec_j|$.
	As a result, according to (\ref{eq:x_u_leftbound}) and (\ref{eq:x_u_rightbound}), one obtains
	\begin{align}\label{eq:med_bound}
	\min \left\{ [\xi^o_i]_j, i\in\Ec_j \right\}\leq \tilde{x}_j\leq \max \left\{ [\xi^o_i]_j, i\in\Ec_j \right\},\ j\in\Uc .
	\end{align}
	Consider the following optimization problem where observation are not influenced by attack:
	\begin{align*}
	\underset{{\tilde{x}}^o,\mu^o}{\text{minimize}}&\quad \frac{1}{2} 
	\begin{bmatrix}
	\mu^o \\
	N H \tilde{x}^o
	\end{bmatrix}^{'} \Wc
	\begin{bmatrix}
	\mu^o \\
	N H \tilde{x}^o
	\end{bmatrix} + \gamma^o\left\|{Y}^o-\mu^o-H \tilde{x}^o\right\|_1 ,
	\end{align*}
	where ${Y}^o$ is composed of $P_i\zeta_i^o$. Denote the solution to this problem as ${\tilde{x}}^o,\mu^o$.
	According to Theorem \ref{th:no_attack}, by choosing 
	$$\gamma^o=	\left\|\Wc \begin{bmatrix}
	\left(I-GF\right)\epsilon^o(k) \\
	N H \hat{x}^o(k)
	\end{bmatrix}\right\|_\ift, $$
	the solution coincides with Kalman estimation, i.e., $\tilde{x}^o(k)=\hat{x}^o(k)$.
	Similar to previous analysis, the solution $\tilde{x}^o$ satisfies 
	\begin{align}\label{eq:normal_solution}
	[\tilde{x}^o]_j=f_{(|\Ec_j|+1)/2}\left( [P_i\zeta_{i}^o-\mu^o_i]_j, i\in\Ec_j \right), \forall j\in\Uc\cup \Sc.
	\end{align}
	Combining \eqref{eq:med_bound} and \eqref{eq:normal_solution} leads to that, for every $j\in\Uc$,
	\begin{align*}
	\left|[\tilde{x}]_j-[\tilde{x}^o]_j\right| =\left|[\tilde{x}]_j-[\hat{x}^o]_j\right|
	\leq &
	\max_{i_1,i_2\in \Ec_j} \left| \left[P_{i_1} \zeta^o_{i_1}(k)\right]_j- \left[P_{i_2} \zeta^o_{i_2}(k)\right]_j \right| 
	+ \|\mu^*\|_\infty+ \|\mu^o\|_\infty\\
	\leq& \max_{i_1,i_2\in \Ec_j} \left| \left[P_{i_1} \zeta^o_{i_1}(k)\right]_j- \left[P_{i_2} \zeta^o_{i_2}(k)\right]_j \right| +(\gamma+\gamma^o)\|\Fc\|_\ift.
	\end{align*}
	Recall that $P_i\epsilon_{i}(k)=P_i \zeta_i (k)-H_ix(k)$. Since for all $i_1,i_2\in \Ec_j$, one obtains $[H_{i_1}x(k)]_j=[x(k)]_j=[H_{i_2}x(k)]_j,\forall k\in\Zb^+$. Thus, we have
	\begin{align*}
	&\max_{i_1,i_2\in \Ec_j} \left| \left[P_{i_1} \zeta^o_{i_1}(k)\right]_j- \left[P_{i_2} \zeta^o_{i_2}(k)\right]_j \right| =\max_{i_1,i_2\in \Ec_j} \left| \left[P_{i_1} \epsilon^o_{i_1}(k)\right]_j- \left[P_{i_2} \epsilon^o_{i_2}(k)\right]_j \right| ,
	\end{align*}
	whose variance is uniformly bounded for all $k$ according to Lemma \ref{lm:epsilon}. Similarly, $\gamma^o$ is also uniformly bounded for all $k$. 
	Recalling that for $j\in\Sc$, we have $\left|\tilde{x}_j(k)\right|\leq\gamma \cdot \left\|\Fc\right\|_\ift$ from \eqref{eq:mu_xs_bounded}. Since $\hat{x}^o_j(k)$ is the oracle Kalman estimation of a stable state, its variance is always bounded. As a result, our estimation $\tilde{x}(k)$ is secure according to Definition \ref{def:resi}. 
\end{proof}

%\nocite{*}% Show all bib entries - both cited and uncited; comment this line to view only cited bib entries;
\bibliographystyle{unsrt}
\bibliography{ref_zishuo}

\clearpage

%\section*{Author Biography}
%
%\begin{biography}{\includegraphics[width=76pt,height=86pt]{lizishuo.jpg}}{\textbf{Zishuo Li} received the Bachelor of Engineering degree in 2019 from School of Automation Science and Electrical Engineering,
%		Beihang University, Beijing, China. He is now working toward the Ph.D. degree in the Department of Automation, Tsinghua University. His research interests include secure control systems, networked control systems, robust estimation and control.}
%\end{biography}
%\vspace{30pt}
%\begin{biography}{\includegraphics[width=76pt,height=86pt]{yilinmo.jpg}}{\textbf{Yilin Mo} is an Associate Professor in the Department of Automation, Tsinghua University. He received his Ph.D. In Electrical and Computer Engineering from Carnegie Mellon University in 2012 and his Bachelor of Engineering degree from Department of Automation, Tsinghua University in 2007. Prior to his current position, he was a postdoctoral scholar at Carnegie Mellon University in 2013 and California Institute of Technology from 2013 to 2015. He held an assistant professor position in the School of Electrical and Electronic Engineering at Nanyang Technological University from 2015 to 2018. 
%		His research interests include secure control systems and networked control systems, with applications in sensor networks and power grids. }
%\end{biography}

\end{document}